\documentclass{article}
\usepackage{graphicx} 
\usepackage{siunitx} 
\DeclareSIUnit\molar{M} 
\DeclareSIUnit\rpm{RPM}
\DeclareSIUnit\g{g}
\DeclareSIUnit\angstrom{\text{Å}}
\usepackage[style=apa]{biblatex} 
\let\cite\parencite 
\usepackage[colorlinks=true, urlcolor=blue, citecolor=blue]{hyperref} 
\bibliography{References}

\newcommand{\delcan}{\delta_\text{can}}
\newcommand{\delflex}{\delta_\text{flex}}
\newcommand{\delz}{\Delta Z}
\newcommand{\kflex}{k_\text{flex}}
\newcommand{\kcan}{k_\text{can}}

\usepackage{authblk}
\title{Experimental methods to control pinned and coupled actomyosin contraction events}
\author[1]{Hyunjae Lee}
\author[1]{James Clarke}
\author[1]{Kyla Wong}
\author[1]{Julia Glenn}
\author[1]{Aniket Marne}
\author[2,3]{Yoichi Miyahara}
\author[1]{Jos\'e Alvarado}
\affil[1]{Department of Physics, University of Texas at Austin}
\affil[2]{Department of Physics, Texas State University}
\affil[3]{Materials Science, Engineering and Commercialization program, Texas State University}

\begin{document}

\maketitle

\begin{abstract}
    Actin and myosin drive many instances of force generation, deformation, and shape change in cells, tissues, and organisms. In particular, cytoskeletal actomyosin is remarkable in its adaptive architecture, responding to a host of actin-binding proteins. Equally important, however, is actomyosin's interaction with its mechanical environment. Actomyosin contractility and environmental properties, such as geometry and stiffness, are inherently coupled. To understand this coupling, novel experimental techniques are needed. Here we describe methods to spatially control the anchoring of reconstituted contractile actomyosin networks to two, opposing surfaces (“transverse anchoring”). The two surfaces can be either rigid (“pinned contraction”), or one of the surfaces may be compliant (“coupled contraction”). We introduce compliance by manufacturing flexure hinges, and describe their calibration. Calibration permits a direct measurement of the contractile force and mechanical work that actomyosin exerts on the environment. The methods described here provide an avenue toward a more complete characterization of actomyosin's role as an actuator, an essential property in its context of driving deformation and shape change in living systems. 
\end{abstract}

\section{Introduction}

Actomyosin contractility powers many mechanical tasks in individual cells, including intracellular transport \cite{khaitlina_intracellular_2014}, cell motility \cite{blanchoin_actin_2014, pollard_cellular_2003, svitkina_actin_2018, paluch_role_2013}, cell division \cite{paluch_dynamic_2006, stachowiak_mechanism_2014, sedzinski_polar_2011}.
Additionally, collectives of cells coordinate contractility to give rise to larger-scale forces and shape-change events.
This coordination is evident in tasks such as morphogenesis and wound healing.
During wound healing, epithelial cells at the wound edge collectively migrate to close the gap, driven by actin-based protrusions and actomyosin contractility that coordinate traction, retraction, and force transmission across the tissue.
Lamellipodia, the leading edge of actin protrusions that propagate cell crawling \cite{pollard_cellular_2003, svitkina_actin_2018}, exert tangential traction forces on the extracellular matrix through focal adhesions, pulling cells while myosin-driven contraction in the lamella and cell rear generates axial tension for retraction \cite{pandya_actomyosin_2017, brugues_forces_2014, vedula_mechanics_2015}.
This maintains the pace of collective migration of the cellular sheet, as the contracting actomyosin network in cell bodies transmits axial tensile forces to rearward through integrin adhesions, enabling follower cells to experience propagated axial stresses for collective advancement. \cite{pandya_actomyosin_2017, svitkina_actin_2018}.
Actomyosin networks also drive shape changes and tissue remodeling during development, playing a big role in tissue morphogenesis.

Actin and myosin are essential components of cellular mechanical function.
Actin is a highly conserved, globular protein (G-actin) that exists ubiquitously in eukaryotic cells, serving as one of the major constituents of the cytoskeleton \cite{pollard_actin_2016}.
It polymerizes reversibly into double-helical filamentous actin (F-actin), forming flexible, polar structures that provide mechanical support \cite{dominguez_actin_2011}.
In conjunction with myosin, an actin-binding motor protein that pulls on actin filaments through ATP-hydrolysis \cite{wang_active_2012}, actomyosin makes up the structural and functional foundation of both sarcomeres and non-muscle cells and generates contractile forces necessary for biological processes.
The contractility of actomyosin networks is structured by its architecture, incorporating both the characteristics of the force-generating myosin fibers and the organization of the actin filaments \cite{koenderink_architecture_2018}.
A diverse set of actin-binding proteins help determine actin architecture and therefore function.
Actin-binding proteins bind two actin filaments together to form crosslinks within the network, affecting the network’s morphology and mechanics \cite{Lieleg-2010-SoftMatter}
However, actin-binding proteins also couple actomyosin to the mechanical environment via adhesion complexes.
These complexes often undergo force-induced conformational changes to expose cryptic binding sites and reinforce cytoskeletal connections \cite{cronin_dynamics_2021}.
Integrins, the anchoring proteins at cell–matrix contacts \cite{zaidel-bar_functional_2007, kanchanawong_nanoscale_2010}, and cadherins, which localize to cell–cell junctions, establish the physical interface between the cytoskeleton and the extracellular environment or neighboring cells.
Meanwhile, talin and kindlin directly bridge integrin cytoplasmic tails to F-actin while transmitting myosin-generated tensions that drive adhesion maturation \cite{roca-cusachs_integrin-dependent_2013}.
Integrin-based focal adhesions serve as key sites for transmitting actomyosin-generated contractile forces to the extracellular matrix, enabling the cell to exert traction and reposition itself \cite{blanchoin_actin_2014}.
In contrast, cadherin-mediated junctions mechanically integrate adjacent cells via the catenin complex and vinculin recruitment under tension, facilitating collective motility by providing directional cues to migrating cells \cite{pandya_actomyosin_2017} and coordinating lamellipodial protrusions in leader cells\cite{gupta_how_2021}.
Focal adhesions also serve as anchors in tissue morphogenesis, where force-dependent mechanical signals are converted into biochemical cues \cite{kasza_dynamics_2011}.
Actomyosin exhibits rich mechanical interactions with adhesions and the surrounding environment.
Contractions can manifest isotropically, where contractile stresses are approximately equal in all directions.
Alternatively, anisotropic contractions can occur along a specified axis, and may be uniform throughout the entire cell body or spatially localized \cite{murrell_forcing_2015}.
Studying actomyosin-environment interactions is therefore important to understand contractility in the context of biological mechanical function.

One key element in actomyosin-environment interactions is the notion of mechanical reciprocity.
As cells contract, they exert forces on their environment and simultaneously experience reciprocal forces in return.
This bidirectional mechanical coupling requires that cells interact with substrates or neighboring structures at specific points along their boundaries, providing anchoring sites for force transmission.
Such boundaries typically occur at interfaces with adjacent cells \cite{wang_active_2012} or the extracellular matrix, but also exist internally, where cytoskeletal elements interact with organelles and membranes, creating a hierarchy of mechanical constraints.
The reciprocal forces imposed by these boundaries lead to the buildup of stress and strain within the actomyosin network.
This accumulation is not static; it drives stress relaxation dynamics that reorganize cytoskeletal architecture, regulate adhesion stability, and modulate force generation over time.
Such feedback mechanisms enable cells to adapt their contractile behavior to changing mechanical conditions, reinforcing adhesions under tension or remodeling them when forces dissipate.
Mechanical reciprocity and feedback couple cellular contractility to environmental stiffness and geometry, setting the stage for complex mechanosensory responses \cite{lappalainen_biochemical_2022}.

The importance of mechanical reciprocity is evident from early studies on cultured cells, which demonstrated that cells generate substantially greater traction forces when adhered to stiffer substrates.
For example, fibroblasts on rigid gels exert higher stresses than on compliant ones, indicating that substrate rigidity enhances force transmission efficiency \cite{lo_cell_2000}.
Local force transmission correlates with focal adhesion size and intensity, with a relatively constant stress per adhesion site \cite{balaban_force_2001}.
Focal adhesion dynamics further stabilize adhesions in a force-dependent manner \cite{zhou_effects_2017} suggesting that both force generation and adhesion reinforcement are modulated by the stiffness of the substrate.
In addition to its effects on focal adhesions, substrate stiffness regulates lamellipodial dynamics and cell motility rates \cite{pelham_cell_1997}, adding cell motility to the range of cellular behaviors shaped by substrate stiffness.
Larger-scale mechanosensing mechanisms have also been observed, implicating cytoskeletal feedback across the cell \cite{trichet_evidence_2012}.
Theoretical models support this view, showing that optimal stiffness for maximal force transmission reflects a balance between molecular dynamics and motor activity \cite{bangasser_determinants_2013}.
Experimental evidence from actomyosin cortical “cytoquakes” further indicates that tension, force fluctuations, and energy dissipation are all regulated by boundary rigidity \cite{shi_pervasive_2021}.
In addition to governing force-generating contractile dynamics, substrate stiffness also acts as a potent mechanosensory cue for cell differentiation and morphogenesis.
For example, mesenchymal stem cells cultured on soft, brain-like matrices adopt neuronal fates and muscle-like and bone-like matrices promote myogenic and osteogenic differentiation, respectively \cite{smith_stem_2018}.
Matrix-elasticity sensing that directs cell fate has specifically been shown to be dependent on the contraction of the actomyosin network \cite{mcbeath_cell_2004}, and as focal adhesion-mediated force transmission enable cellular mechanosensing of the microenvironment \cite{engler_matrix_2006}.
Recent studies suggest that cells may sense strain energy rather than stiffness per se \cite{panzetta_cell_2019}, highlighting the significance of a more complex dynamic in cell–boundary interactions than implied by elastic modulus alone.
In sum, understanding contractile mechanics under varying boundary conditions is therefore essential for elucidating their dual role in regulating cellular force generation and energy transfer, as well as serving as a key environmental signal that directs cell behavior and fate.

While the molecular mechanisms underlying actin polymerization and actomyosin contraction, such as filament assembly, ATP-driven motor activity, and crosslinking, are increasingly well understood \cite{thoresen_reconstitution_2011}, knowledge gaps remain regarding the mechanical properties of actomyosin interactions during axial force transmission and stress propagation of non-sarcomeric actin architectures \cite{murrell_f-actin_2012}.
Current experimental approaches have shed light on actomyosin mechanics.
In vitro systems have been instrumental in dissecting these mechanics under controlled conditions.
By reconstituting purified actin filaments and myosin motors in simplified environments, these systems isolate fundamental principles of contractility without the complexity of living cells.
Compared to studies in vivo, they lack the full complement of regulatory proteins, signaling pathways, and dynamic interactions with membranes and organelles.
However, this reductionist approach offers significant advantages, enabling precise control over network composition, geometry, and boundary conditions, and facilitating direct visualization and manipulation of actomyosin dynamics.
Such systems provide a powerful platform for testing theoretical models and uncovering emergent behaviors that might be obscured in the crowded intracellular environment.
At the same time, translating insights from these simplified systems to cellular contexts requires careful consideration of additional layers of regulation and mechanical feedback present in vivo.
Free actomyosin contraction assays reconstitute gels of actin filaments and myosin motors without fixed boundaries.
Unanchored to any boundaries and allowed to contract without external constraints, free contraction models isotropic contractions, as the forces generated by myosin activity lead to uniform shrinkage of the gel rather than directional tension.
These assays are valuable for studying how intrinsic properties such as filament density, crosslinking, and network architecture govern contractile behavior, independent of external mechanical cues \cite{alvarado_molecular_2013, alvarado_force_2017, reymann_actin_2012, ideses_spontaneous_2018, livne_self-assembled_2024}.
However, they differ from cellular contexts, where actomyosin networks interact with membranes and extracellular matrices, introducing boundary conditions and feedback that strongly influence contraction patterns.
Among studies that do impose boundary conditions, most exhibit lateral anchoring, where a planar gel is tethered parallel to its surface.
Although this may serve as a model the actomyosin cortex, which is anchored to the inner leaflet of the plasma membrane, transverse anchoring, which permits larger-scale rearrangements of the gel and accumulation of stress, has not been explored as extensively.
A widely utilized experimental approach that incorporates adhesion to a boundary condition is traction force microscopy.
Traction force microscopy (TFM) quantifies the forces exerted by cells on elastic substrates by tracking the displacement of embedded fluorescent fiducial markers as cells pull on the surface of the substrate \cite{style_traction_2014} and has been pivotal for analyzing cellular generation and transmission of traction forces during migration, adhesion, and mechanotransduction, \cite{dembo_stresses_1999, banda_reference-free_2019, grashoff_measuring_2010, kollimada_biochemical_2021, mohl_quantitative_2012, plotnikov_force_2012, kraning-rush_role_2011} and is increasingly applied to multicellular systems at molecular, tissue, and organismal scales \cite{style_traction_2014}.
TFM assays have also been used to measure strain energy and contractile work in fibroblasts \cite{de_la_pena_quantifying_2021} and myofibroblasts \cite{hinz_quantitative_2021}.
This better mimics stress propagation in embryos and tissues, where a supracellular, three-dimensional contractile network connected by cell-cell and cell-matrix adhesions accumulates stress and effects global shape change.
Rheological methods have also characterized the mechanical properties of actomyosin networks, revealing how myosin modulates elasticity and stiffness in response to internally generated contractile stress \cite{koenderink_active_2009, hale_resolving_2009}.
There have been studies that anchor actomyosin gels transversely, for example by attaching deformable oil droplets to opposite ends of a gel \cite{bendix_quantitative_2008}.
However, there are limitations of such boundary controls, as oil droplets will rupture before maximal stress loading of the contractile gel and only provide two deformable boundaries, which warrants a need for methods for an extensive study of contractile actomyosin gels under transverse anchoring.
Despite their utility, these established assays and models for contractile gels have notable limitations.
Unconstrained free-contraction assays lack biologically relevant boundary conditions, TFM primarily models tangential force application at interfaces, and conventional rheological methods focus on shear stresses, all which does not capture the transverse boundary interactions that also arise in cellular environments.

Here we present experimental methods for spatial control over boundary adhesion of reconstituted actomyosin active gels.
We adapt protocols for building chambers where entire gels can be observed in a fluorescence microscope \cite{alvarado_reconstituting_2015}.
We further adapt an existing protocol to selectively nucleate actin \cite{reymann_nucleation_2010}, thereby pinning actomyosin gels to two opposing boundaries (“transverse anchoring”).
Finally, we implement flexure hinges, an elastically deformable beam with linear spring constant \cite{lobontiu_compliant_2020}.
We adhere contractile actomyosin to one rigid surface and one flexure hinge.
The gel deforms the hinge, providing a measure of the force exerted and the work performed by the gel on the hinge.
These methods will enable detailed investigations of crucial network mechanics such as stress relaxation, strain formation, and force generation under transverse anchoring that replicate aspects of mechanical environments during cell motility, morphogenesis.
Furthermore, these systems may facilitate the exploration of how actomyosin networks respond to external stresses, integrate contractile forces across multiple boundary conditions, and how network architecture governs force transmission and mechanical stability.
Insights gained will deepen understanding of fundamental processes such as tissue morphogenesis, cellular mechanotransduction pathways, and cytoskeletal organization principles, and ultimately, we hope this approach can help bridge the gap between molecular-scale actomyosin dynamics and biologically relevant tissue-scale biomechanics.

\section{Methods}

This methods section outlines the experimental procedures used to construct and prepare sample chambers for in vitro actomyosin contraction assays. 
The process begins with the cleaning and preparation of glass slides and cover slips to create hydrophilic, contaminant-free surfaces. 
We assemble these cleaned surfaces into chambers using specific components that define two specific chamber geometries: a pinned chamber with rigid lateral boundaries we call “wings” (Fig. \ref{fig:pinned}) and a flexure chamber incorporating an acrylic flexure hinge that allows controlled mechanical deformation in response to contractile forces (Fig. \ref{fig:flexure}).
The surfaces of the chambers are then functionalized through chemical passivation and selective activation steps. 
Protein reagents are prepared with high-speed centrifugation, quantified via absorbance spectroscopy, and combined using standard wet-lab techniques according to precise stoichiometric ratios. 
The resulting actomyosin mixtures are introduced into the prepared chambers to study contraction under controlled conditions.
Epifluorescence imaging is performed for time-lapse imaging that captures the progression of actin polymerization and network contraction, with DAPI illumination used to trigger ATP release and initiate active contraction events. 
Together, these methods enable reproducible visualization of actomyosin-driven mechanics at well-defined boundaries.

\subsection{Chamber components}

In order to assemble pinned and flexure chambers, we first prepare the various components needed. 
A glass slide is used as the base of the chamber to which we attach the “wing” lateral boundaries, and a cover slip is used to cover the chamber from the top. 

We first clean chamber components by adapting a previous protocol based on base piranha \cite{alvarado_reconstituting_2015}. The glass slides (VWR VistVision: 75x25x1 \unit{\milli\meter}) and cover slips (VWR Microscope Cover Glasses: 24x60 \unit{\milli\meter}) are first cleaned by sonication in deionized water for five minutes, followed by immersion in a base piranha solution composed of five parts deionized water, one part \textgreater 30\% ammonium hydroxide, and one part \textgreater 25\% hydrogen peroxide at \qty{80}{\celsius} for \qty{30}{\minute}. 
This removes trace organic and carbon residues and makes the glass highly hydrophilic \cite{schmidt_safe_2022}, ensuring contaminant-free surfaces and improving subsequent surface passivation and functionalization treatments. 
After the piranha solution, we sonicate the slides and cover slips in deionized water for \qty{5}{\minute}, and then blow dry and store them in an isopropanol solution. 
The cover slips are cut to the desired size that covers the entire chamber area by scoring them with a glass cutter and breaking them along the scores.
Parafilm is pressed onto glass slides, and a design with the specified dimensions of Parafilm layers needed is printed onto the glass slide with a laser cutter. 
The excess Parafilm is carefully removed to not damage the Parafilm needed for the chamber. 

\begin{figure}
    \centering
    \includegraphics[width=0.5\linewidth]{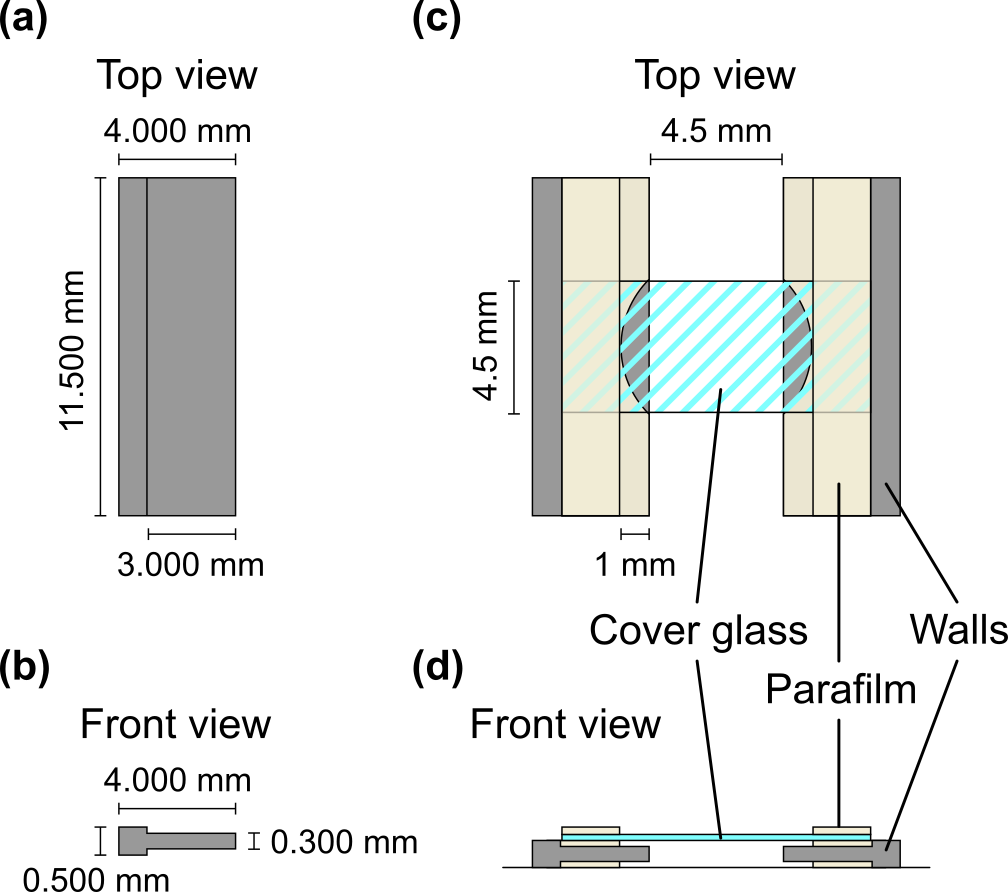}
    \caption{Pinned contraction chamber \\ A schematic of the sample chamber used in pinned contraction assays. \\ (a) Top view of the chamber wings constructed from 316 L stainless steel. \\ (b) Front view of the chamber wings. Depicted with 2:1 aspect ratio (height:width) for clarity. \\ (c) Top view of the assembled pinned contraction chamber, with steel wings (dark gray) Parafilm (yellow) and cover glass (blue). \\ (d) Front view of the assembled chamber. Depicted with 2:1 aspect ratio (height:width) for clarity.}
    \label{fig:pinned}
\end{figure}

The wings that make up the fixed lateral boundaries of the chamber (Fig.\ref{fig:pinned}a, \ref{fig:pinned}b) are designed using AutoCAD and made with 316L stainless steel.
These serve as actin nucleation sites in both the pinned chambers and the flexure chambers and seal the lateral edges of the chamber. 
The ledges are designed to align with a layer of Parafilm under the wings that secures the wings to the glass slide and a layer of Parafilm on top of the wings that secures the cover slip to the wings. 
The segment of the top and bottom face of the wings exposed to the interior of the chamber is the actin nucleation site. 

\subsection{Pinned chamber}
We use the cleaned glass slide, cover slip, and steel wings to assemble the pinned chamber (Fig. \ref{fig:pinned}c and \ref{fig:pinned}d).
The wings lay on both sides of the chamber, facing each other, on top of a layer of Parafilm. 
We lay a matching layer of Parafilm on top of and under the wings, with a curved cutout that exposes an area of the wing for the actin nucleation site. 
The cutout is curved to facilitate the ejection of liquids during the blow drying of the passivation and activation steps. 
We attach a cover slip cut to the desired dimensions (\qty{4.5}{\milli\meter} x \qty{10.5}{\milli\meter}) to the layer of Parafilm on top of the wing, and an additional layer of Parafilm is added on the outer edges to secure the cover slip. 
The chamber is heated until the Parafilm turns clear and is pressed down to secure the chamber. 
Heated Parafilm turns pliable and adhesive, and thus can be used to attach the components of the chamber together while creating a seal. 
After the actomyosin assay is pipetted into the chamber, the upper and lower openings are sealed by a layer of vacuum grease. 
The sealing provided by the Parafilm and vacuum grease isolate the chamber during experimentation, minimizing the effects of outside contamination, leakage, and evaporation during imaging.

\begin{figure}
    \centering
    \includegraphics[width=0.5\linewidth]{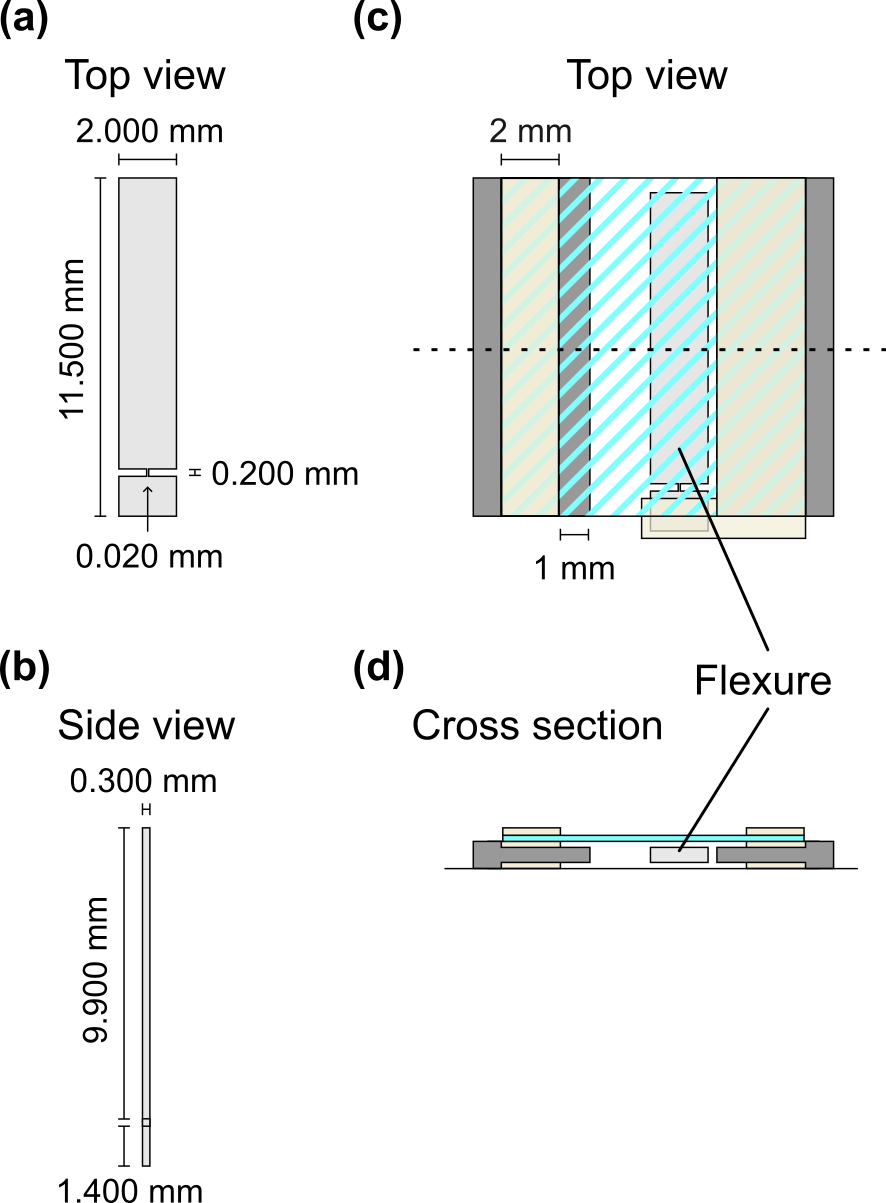}
    \caption{Flexure Chamber \\ A schematic of the sample chamber used in flexure experiment assays. \\ (a) Top view of the flexure, made of acrylic. \\ (b) Side view of the flexure. \\ (c) Top view of the assembled flexure chamber, with  flexure (light gray), steel wings (dark gray), Parafilm (yellow), and cover glass (blue). \\ (d) Transverse cross section of the flexure chamber (cf. panel c, dashed line). Depicted with 2:1 aspect ratio (height:width) for clarity.}
    \label{fig:flexure}
\end{figure}

\subsection{Flexure chamber}
A flexure, or flexure hinge, is a specially designed flexible structure that allows controlled bending or elastic deformation, enabling precise motion in specific directions while restricting movement in others \cite{lobontiu_compliant_2020}. 
The elastically deformable element is often connected to rigid bodies; some of these bodies are anchored and the other bodies move relative to the anchoring as the flexure bends. 
For single-axis flexures, this bending occurs two-dimensionally, although multi-axis flexures can bend about multiple compliant axes called the bending-sensitive axis.

We design the experimental flexure (Fig. \ref{fig:flexure}a and \ref{fig:flexure}b) using detasFLEX software and use polymethyl methacrylate (PMMA) as the material with the given dimensions to best match the range of forces expected by actomyosin contraction. 
The single-axis flexure has two rigid bodies, a smaller base that anchors to the glass slide, and a longer arm free to bend along its bending-sensitive axis.
Due to the greater thickness (\qty{0.300}{\milli\meter}) of the flexure compared to the width (\qty{0.020}{\milli\meter}), the flexure bends only slide-to-side around the bending-sensitive axis, which is normal to both the length and width of the flexure,  with roughly \qty{11}{\degree} of maximum amplitude.
This bending occurs in response to a net torque about its bending-sensitive axis, and the compliance or stiffness of the flexure hinge can be quantified by a spring constant. 
We determine the spring constant as described in Section 3.2.

The design and assembly of the flexure chamber (Figs. \ref{fig:flexure}c and \ref{fig:flexure}d) are slightly different to the pinned chamber. 
We lay the wings vertically on both sides of the chamber; however, the top and bottom Parafilm layer for one wing exposes the wing down its whole length rather than the small curved cutout, while the top and bottom Parafilm layer for the opposite wing covers it down its whole length.  
The entire uncovered area of the exposed wing serves as the actin nucleation site and the fixed boundary. 
The opposite wall fully covered in Parafilm acts as a passive wall to seal the other side of the chamber. 
The flexure is attached on the side of the chamber closer to the fully covered wing, and serves as the other actin nucleation site and the compliant boundary. 
The base is sandwiched between the glass slide and the cover slip by a layer of Parafilm on each side. 
A cover slip of the desired size (\qty{11.5}{\milli\meter} x \qty{10.5}{\milli\meter}) is used to cover the entirety of the area, and a layer of Parafilm on top is used to secure the cover slip. 
To secure the components of the chamber, the Parafilm is heated until clear and then pressed down. 
After the actomyosin assay is pipetted into the chamber, the openings in the top and bottom of the chamber are sealed with a layer of vacuum grease. 

\begin{figure}
    \centering
    \includegraphics[width=0.5\linewidth]{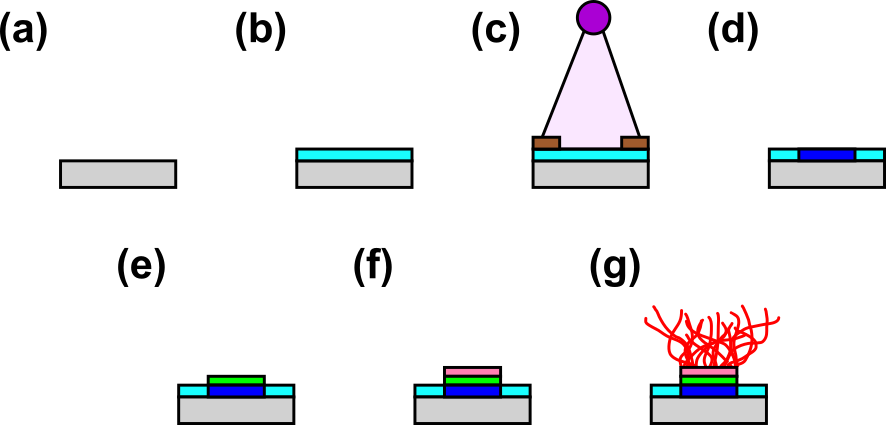}
    \caption{Selective nucleation of actin \\ Steps to initiate actin nucleation at defined adhesion sites. \\ (a) Surface of the sample chamber. \\ (b) KOH-treated surfaces coated with PLL-PEG (light blue). \\(c) Selective exposure of UV (purple) to PLL-PEG of actin adhesion site specified by a mask (brown). \\ (d) Degradation of PLL-PEG passivation (dark blue) at the adhesion site. \\ (e) Attachment of GST-VCA (green) at the adhesion site. \\ (f) Interaction of Arp2/3 (pink) in protein assay with GST-VCA on the adhesion site. \\ (g) Actin nucleation (red) at the adhesion site.}
    \label{fig:nucleation}
\end{figure}

\subsection{Passivation and activation}
Passivation and activation are critical surface treatment steps that are used to prepare experimental chambers for selective biochemical interactions that enable spatial control of adhesion and nucleation sites for the actomyosin gel.

The sample chambers are rinsed alternately between ethanol (EtOH) and Milli-Q using nitrogen gas to blow dry in between each. 
During this step, we estimate the chamber volume which dictates the volume of reagents we prepare for incubation solutions and the final actomyosin solution. 
This is done until no air bubbles are seen as the liquids are pipetted into the chamber, and we do a final wash with Milli-Q. 
The chambers are then slightly overfilled with KOH (\qty{1.0}{\molar}) and set into a Petri dish with a Kimwipe oversaturated with Milli-Q for \qty{10}{\minute} (Fig. \ref{fig:nucleation}a). 
The chamber is overfilled with solution to ensure complete coverage of all surfaces and to account for possible evaporation, and the oversaturated Kimwipe is used to maintain a humid environment inside the Petri dish, reducing evaporation of solution in the chamber. 
Using nitrogen gas, the chamber is cleared from KOH and rinsed with Milli-Q. 
These steps rinsed out organic substrates that could cause non-specific binding, as well as modifying the glass surface to be more negatively charged. 
This negative charge allows for adsorption of the positively charged PLL polyelectrolytes to the chamber surfaces.
Adsorption via electrostatic interactions is strongest for materials with a certain density of charges such as glass and steel \cite{Movilli-2020-Org.Mater.}.
PLL-PEG (\qty{0.2}{\milli\gram\per\milli\liter} of Milli-Q) is added to slightly overfill the chamber, which is incubated in a Petri dish with a Kimwipe oversaturated with Milli-Q for \qty{45}{\minute} (Fig. \ref{fig:nucleation}b). 
Finally, the chambers are rinsed with Milli-Q and dried with nitrogen gas. 
The surfaces are coated with PLL-PEG which passivate the surfaces through steric shielding and electrostatic absorption of PEG. 
This prevents any unwanted interactions between any of the surfaces not meant to interact with the other reagents that we flow into the chamber later.  

After the attachment of PLL-PEG, we place a mask (laser printed from chipboard) to isolate the desired actin nucleation sites of the chamber and expose them to UV light (\qty{254}{\nano\meter}) for 5 minutes on both sides of the glass slide (Fig. \ref{fig:nucleation}c). 
The curved cutouts on both wings are the actin nucleation sites for the pinned chamber, and the entire exposed edge as well as the entire arm of the flexure are the actin nucleation sites for the flexure chamber. 
The UV exposure denatures PLL-PEG and opens up reactive groups on the PLL-PEG surface, allowing for covalent linkages, electrostatic binding, or strong non-covalent interactions (Fig. \ref{fig:nucleation}d) \cite{arima_complement_2008}. 
After exposure, the chamber is overfilled with GST-VCA solution (\qty{3}{\micro\molar}) for \qty{15}{\minute} in a Petri dish with a Kimwipe oversaturated with Milli-Q (Fig. \ref{fig:nucleation}e). 
This binds GST-VCA to the selected actin nucleation sites, which with the Arp 2/3 protein complex in the actomyosin assay serve as the actin nucleation promoting factor (NPF) that selectively nucleates and attaches actin to the specific sites that GST-VCA is attached (Fig. \ref{fig:nucleation}f). 
Profilin is added to hinder free nucleation of actin filaments, directing nucleation to start at the desired attachment sites, and also promoting continued polymerization of actin networks from the nucleation sites (Fig. \ref{fig:nucleation}g) \cite{blanchoin_actin_2014}. 

\subsection{Reagent preparation}
The experiment requires proteins including actin, actin labeled with Alexa Fluor 594, myosin, fascin, Arp2/3, profilin, and GST-VCA, as well as other reagents that need to be prepared to execute. 
We now turn to the preparation needed for the reagents. 

We purify monomeric (G-) actin and myosin II from rabbit psoas skeletal muscle \cite{alvarado_reconstituting_2015}, which yields unlabeled actin and actin labeled with Alexa Fluor 594 carboxylic acid, succinimidyl ester \cite{soares_e_silva_active_2011}.
Fascin is purified using T7 \textit{Escherichia coli} and GST-fascin pGEX vector \cite{alvarado_reconstituting_2015}. 
Arp2/3 protein complex from porcine brain is obtained from Cytoskeleton (RP01P) and reconstituted with distilled water to a buffer of 20 mM Tris pH 7.5, 25 mM KCl, 1 mM MgCl2, 0.5 mM EDTA, 0.1 mM ATP, 1\% dextran and 5\% sucrose and stored at \qty{-80}{\celsius}. 
Both human recombinant profilin 1 from Cytoskeleton (PR02) and the fragment recombinant protein of human profilin 1 from ThermoFisher (RP-108174) are used, with the Cytoskeleton profilin stored in \qty{10}{\milli\molar} Tris pH 8.0, \qty{1}{\milli\molar} EDTA, \qty{1}{\milli\molar} DTT, 5\% (w/v) sucrose and 1\% (w/v) dextran and the ThermoFisher profilin is stored in \qty{1}{\molar} urea/PBS, pH 7.4, both at \qty{-80}{\celsius}. 
GST-VCA is obtained from Cytoskeleton as well (VCG03), and reconstituted in \qty{20}{\milli\molar} Tris pH 7.5, \qty{25}{\milli\molar} KCl, \qty{1}{\milli\molar} MgCl2, \qty{0.5}{\milli\molar} EDTA, 0.2\% dextran, 2\% sucrose and stored at \qty{-80}{\celsius}. 

The day before use, we dialyze the stock solutions of actin and Arp2/3 overnight at \qty{4}{\celsius} in the following buffer: \qty{5}{\milli\molar} Tris-HCl pH 7.8, \qty{0.1}{\milli\molar} calcium chloride, \qty{5}{\milli\molar} dithiothreitol, \qty{0.2}{\micro\molar} NaATP pH 7.8. 
Myosin dialysis also happens overnight at \qty{4}{\celsius} in the following buffer: 20 mM imidazole pH 7.4, 300 mM potassium chloride, 4 mM magnesium chloride, 1 mM dithiothreitol \cite{alvarado_reconstituting_2015}. 

On the day of experimentation, we use a centrifuge to spin unlabeled actin, actin labeled with Alexa Fluor 594, myosin, fascin, Arp 2/3, GST-VCA, and profilin  at \qty{120000}{\g}, at \qty{4}{\celsius} for \qty{5}{\minute} to remove aggregates.
The supernatants are collected after centrifuging. 

The absorbance of the supernatants are measured using a NanoDrop 2000, and using the molar absorbance coefficients as shown in Table \ref{tab:molar_absorbance}, we calculate the respective concentrations using the Beer-Lambert law.

\begin{table}
    \centering
    \begin{tabular}{|c|c|c|}\hline
         Reagent&  Wavelength (\unit{\nano\meter})& Molar Absorbance Coefficient (\unit{\per\molar\per\centi\meter})\\\hline
         Actin&  290& 26,600\\\hline
         Alexa594&  594& 92,000\\\hline
         Myosin&  280& 249,000\\\hline
         Fascin&  280& 66,280\\\hline
         Arp2/3&  290& 139,000\\\hline
         Profilin&  280& 18,575\\\hline
         GST-VCA&  280& 46,200\\ \hline
    \end{tabular}
    \caption{Molar absorbance coefficient of reagents}
    \label{tab:molar_absorbance}
\end{table}

GST-VCA is diluted with Milli-Q to \qty{3}{\micro\molar} to match the [GST-VCA] to [actin] ratio of \qty{0.5}{\micro\molar}:\qty{2}{\micro\molar} before being flowed into the sample chamber for the \qty{15}{minute} incubation.

The final concentrations of the actomyosin assay are given in Table \ref{tab:actomyosin}, with the final concentration of the master buffer given in Table \ref{tab:master_buffer}.

\begin{table}
    \centering
    \begin{tabular}{|c|c|}\hline
         Reagent& Concentration\\\hline
         Actin& \qty{12}{\micro\molar}\\\hline
         Alexa594& \qty{0.12}{\micro\molar}\\\hline
         Myosin& \qty{0.12}{\micro\molar}\\\hline
         Fascin& \qty{1.2}{\micro\molar}\\\hline
         Arp2/3& \qty{0.18}{\micro\molar}\\\hline
         Profilin& \qty{36}{\micro\molar}\\\hline
         Master Buffer& Table \ref{tab:master_buffer}\\\hline
         Phalloidin& \qty{12}{\micro\molar}\\ \hline
    \end{tabular}
    \caption{Actomyosin assay concentration}
    \label{tab:actomyosin}
\end{table}

\begin{table}
    \centering
    \begin{tabular}{|c|c|}\hline
         Reagent& Concentration\\\hline
         Imidazole pH 7.4& \qty{20}{\milli\molar}\\\hline
         Potassium chloride& \qty {50}{\milli\molar}\\\hline
         Magnesium chloride& \qty{2}{\milli\molar}\\\hline
         Dithiothreitol& \qty{5}{\milli\molar}\\\hline
         Protocatechuic acid& \qty{2}{\milli\molar}\\\hline
         Protocatechuase 3,4-dioxygenase & \qty{0.1}{\micro\molar}\\\hline
         Trolox& \qty{1}{\milli\molar}\\ \hline
 Caged adenosine triphosphate&\qty{0.1}{\milli\molar}\\\hline
    \end{tabular}
    \caption{Master buffer concentration}
    \label{tab:master_buffer}
\end{table}

The labeling ratio is given by the ratio of actin labeled with Alexa594 to total actin concentration, and has ranged from between 0.01 to 0.05 depending on volume restrictions. 
Changes in labeling ratio can be accommodated for by adjusting the brightness and contrast of images. 

When preparing the solution, a specific order is followed to minimize unwanted protein interaction before reaching the sample chamber. 
First, we combine the master buffer of 5x concentration and deionized water in one tube. 
Myosin, caged ATP, fascin, and Arp2/3 profilin are added to this tube and mixed well by pipetting. 
We then combine unlabeled actin and labeled actin  in a separate tube and mix well by pipetting. 
The contents of the first tube are pipetted into the second tube using a \qty{200}{\micro\liter} Eppendorf pipette tip and immediately mixed well by pipetting.
This specific pipette tip size is used because of its slender shape, which allows the user to circulate the tip during aspiration, allowing a more even mixture of the viscous protein solution.
The contents of the second tube are immediately pipetted into the tube prepared with evaporated phalloidin and mixed well by pipetting. 
Lastly, we pipette the contents of this final tube into the sample chamber.

\begin{figure}
    \centering
    \includegraphics[width=0.25\linewidth]{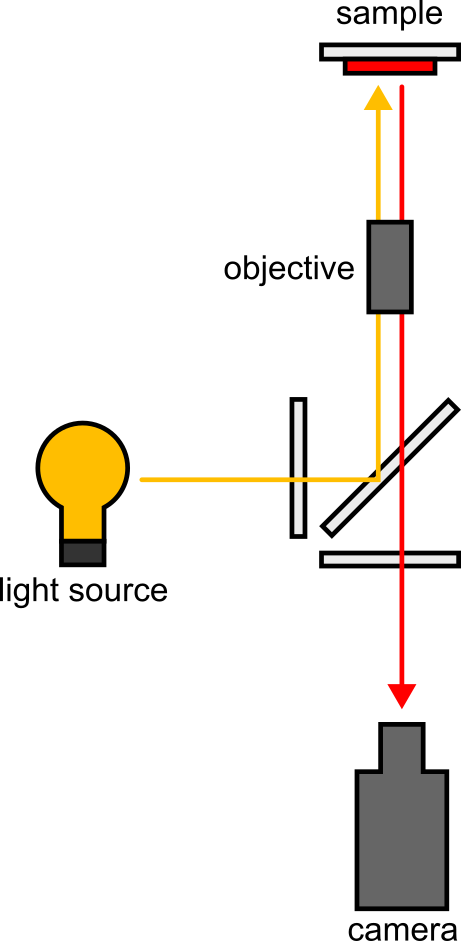}
    \caption{Epifluorescence Microscopy \\ Schematic of the epifluorescence microscopy setup. Excitation light (filtered using an mPlum filter) is directed through the objective onto the sample, where fluorophores absorb the light and re-emit fluorescence. The emitted light is filtered for the desired wavelength before being captured by the camera.}
    \label{fig:microscopy}
\end{figure}

\subsection{Imaging}
Actin polymerizes very soon after components are mixed, and myosin begins to slide actin filaments immediately after they have formed.
Hence, it is important to minimize the time between mixing the sample and starting image acquisition.
To minimize this time, the position of the sample chamber is determined with the microscope software beforehand and the $z$-axis focus is set.
We pipette the freshly mixed actomyosin solution into the sample chamber, and immediately seal the top and bottom openings of the chamber with vacuum grease. 
Even if the chamber is fixed on the microscope stage, its position may change slightly upon loading and sealing.
We therefore adjust the position and $z$-axis focus as needed.

Images are captured every \qty{20}{\second} until there are no visible changes in the gel. 
If the actomyosin gel is not fully contracted, the gel is exposed to \qty{1}{\second} of DAPI to release caged ATP. 
The DAPI exposure time is increased to intervals of \qty{2}{\second}, \qty{5}{\second}, and \qty{10}{\second} depending on the progress of contraction. 

As shown in Fig. \ref{fig:microscopy}, the samples are imaged in a Zeiss Axio Observer 5 epifluoresence microscope with a broad spectrum Lumen Dynamics X-Cite Xylis LED.
We image at 60\% intensity and an exposure time of \qty{2.5}{\milli\second} through a EC Plan-NEOFLUAR 2.5x NA 0.085 objective (Zeiss), and images are captured with a Teledyne Photometrics Prime BSI Scientific CMOS camera.
To excite the Alexa594 dye the actin filaments are labeled with, we use the mPlum of the Zeiss 64 HE filter set, which has an excitation window of \qty{570}{\nano\meter} to \qty{600}{\nano\meter} and emission window of \qty{605}{\nano\meter} to \qty{680}{\nano\meter}.

To release caged-ATP, we use DAPI from the Zeiss 49 filter set, which has an excitation window of \qty{300}{\nano\meter} to \qty{395}{\nano\meter}and an emission window of \qty{410}{\nano\meter} to \qty{460}{\nano\meter}.
DAPI intensity is set to 100\% with variable exposure from \qty{0.5}{\second} to \qty{10}{second}.

\section{Analysis}

\begin{figure}
    \centering
    \includegraphics[width=.9\linewidth]{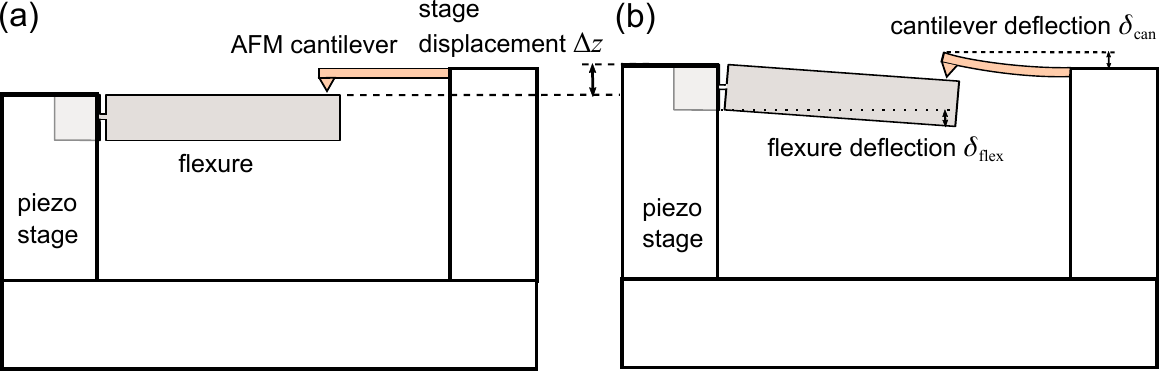}
    \includegraphics[width=1\linewidth]{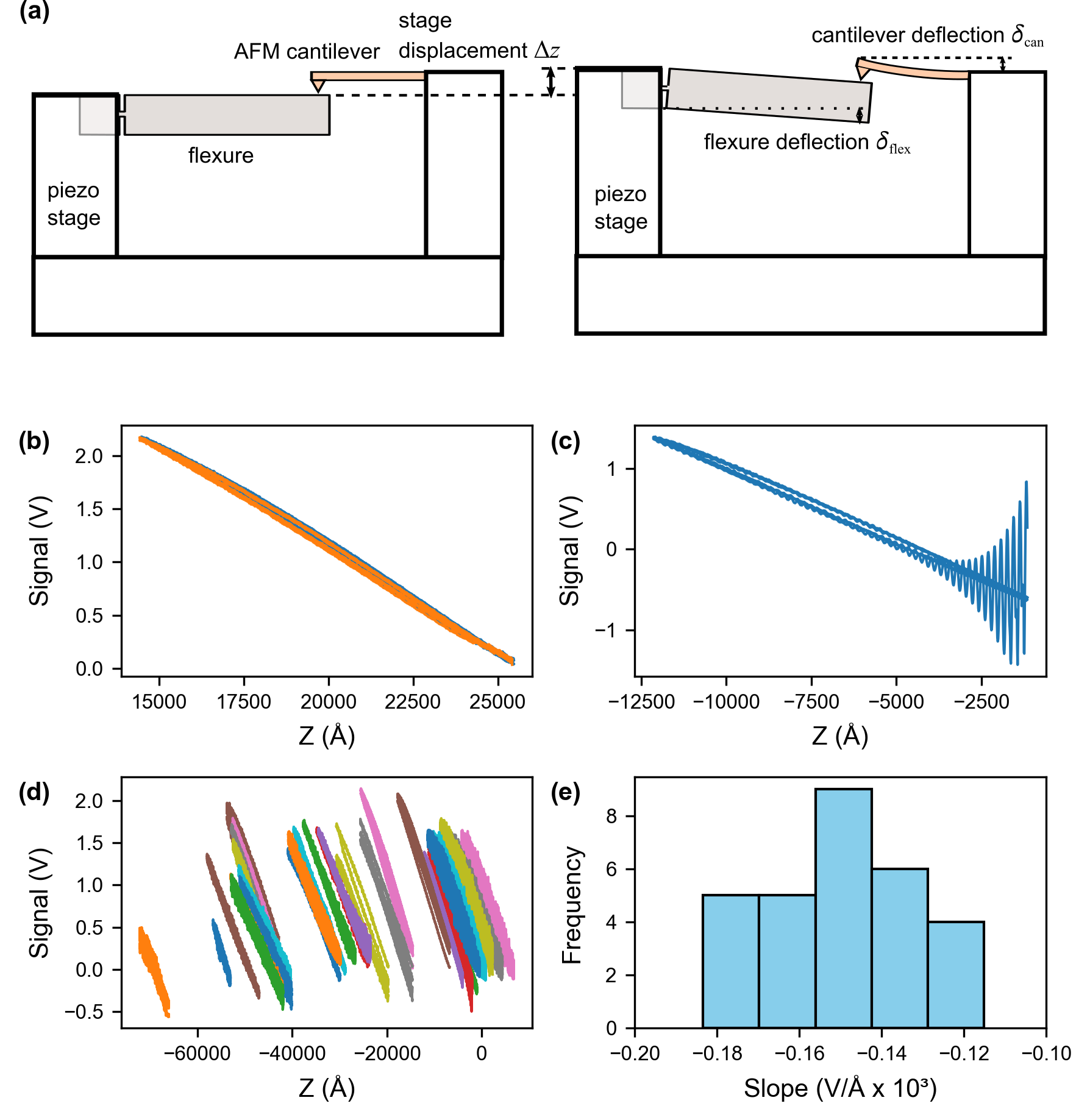}
    \caption{Flexure Calibration \\ The flexure is calibrated using atomic force microscopy (AFM) with a probe of known stiffness to determine the spring constant of the flexure for force calculations. 
    (a) Schematic of calibration setup. 
    (b) Combined reference measurements of $n=2$ datasets of the photodiode signal as a function of Z-piezo displacement in Angstroms. (c) Single flexure measurement dataset of the photodiode voltage signal as a function of Z-piezo displacement in Angstroms. 
    (d) Combined measurements of multiple flexure positions along the same flexure ($n=29$).
    (e) Histogram showing the frequency distribution of fitted slope,
    $S^m$.  $\bar{S}^m$= \qty{-1.47e-04}{\volt\per\angstrom} $\pm$ \qty{5.06e-08}{\volt\per\angstrom}} .
    \label{fig:calibration}
\end{figure}

The flexure experiment was conducted in the flexure chamber, with one side of the gel anchored to a rigid, nondeformable lateral boundary, and the opposite side of the gel anchored to the arm of the compliant, elastic flexure. 
As the actomyosin gel contracts, it pulled the flexure towards the nondeformable boundary. 
By tracking the displacement of the flexure and referencing its calibrated spring constant, a force measurement is calculated for one experiment. 

\subsection{Flexure calibration}
The spring constant of the flexure, $\kflex$, is calibrated using an atomic force microscopy (AFM) cantilever probe
with a known spring constant, $\kcan$,
as a calibration standard (Fig.~\ref{fig:calibration}a)
\cite{torii1996}. 
In AFM, a vertical deflection of an AFM cantilever probe (Fig.~\ref{fig:calibration}a right) is measured by a position-sensitive photodiode.
The laser beam reflected from the cantilever strikes different positions on the photodiode depending on the cantilever deflection.
The resulting photodiode signal, $\Delta V$, is proportional to the vertical deflection of the cantilever, $\delcan$, according to $\Delta V = S^0 \delcan$
where $S^0$ is called the deflection measurement sensitivity. 

When the base of the flexure is moved up by an amount of $\delz$ by a piezo stage,
the deflections of both the flexure, $\delflex$, and the cantilever, $\delcan$, result
(Fig.~\ref{fig:calibration}b).
Considering the relation between two deflections and the displacement of the flexure base, $\delz= \delflex + \delcan$, 
and the force balance between the restoring forces of the flexure and the cantilever, 
$\kflex \delflex = \kcan \delcan$, 
we obtain the following relation, 
\begin{equation}
\delcan = \frac{\kflex}{\kcan + \kflex} \delz.
\end{equation}
The measured photodiode signal, $\Delta V^m$, is given by $\Delta V^m = S^0 \delcan$.
Therefore, the slope of the measured photodiode signal vs the piezo stage displacement, $S^m$ is given as
\begin{equation}
    S^m = \frac{\Delta V^m}{\delz} = \frac{\kflex}{\kcan + \kflex} S^0.
\end{equation}
Solving this equation for $\kflex$ yields
\begin{equation}
\kflex = \kcan \times \frac{r}{1-r}    
\end{equation}
where $r = S^m/S^0$.

In the experiments,  the slope of the photodiode signal vs piezo displacement is first recorded while the AFM probe tip is in contact with a sapphire surface to measure the deflection measurement sensitivity, $S^0$.
In this case, the deflection and deformation of the sapphire substrate is negligibly small (corresponding to $\kflex = \infty$ and thus $\delcan = \delz$). 
An ordinary least squares linear regression on the reference measurement is shown in Fig. \ref{fig:calibration}b. 
Next, the base of the flexure is clamped to a rigid base while the AFM probe is in contact with the arm of the flexure.
As the AFM probe bends the flexure at its hinge, 
the photodiode signal vs the piezo stage position curves are measured (Fig. \ref{fig:calibration}c).
The measurements are repeated for many different contact positions along the flexure (Fig. \ref{fig:calibration}d). 
Before compiling and calculating the sensitivity of the flexure, the data is trimmed at the edges to preserve linearity and reduce noise near the limits.

Across 29 trials, $S^m$ appear roughly normally distributed (Fig. \ref{fig:calibration}e), even when measured at different positions along the flexure.
Because each linear fit carries a different level of uncertainty depending on noise and local measurement conditions, a weighted average of $S^0$ and $S^m$, $\bar{S}_{\rm cantilever}$ and $\bar{S}_{\rm flexure}$ is used instead of a simple mean.
This approach assigns greater weight to slopes with smaller standard errors ensuring that the final comparison to the reference slope, and the resulting spring constant calculation, reflect the most reliable measurements while minimizing the influence of noisier data.

For the model experiment, $\bar{S}_{\rm cantilever} = \qty{-1.97e-4}{\volt\per\angstrom} \pm \qty{1.01e-7}{\volt\per\angstrom}$,  $\kcan$ = \qty{0.141}{\newton\per\meter}, $\bar{S}_{\rm flexure} = \qty{-1.46e-04}{\volt\per\angstrom} \pm \qty{5.06e-08}{\volt\per\angstrom}$, and $\kflex = \qty{4.04e-01}{\newton\per\meter} \pm \qty{1.18e-04}{\newton\per\meter}$.

The AFM cantilever used is CONT-50 cantilevers (Nanosensors)
with the length of 450\,$\mu$m, the width of 50\,$\mu$m and the nominal spring constant of 0.2\,N/m.
We calibrated the spring constant of the cantilever by using Sader's method \cite{Sader99}.

\begin{figure}
    \centering
    \includegraphics[width=1\linewidth]{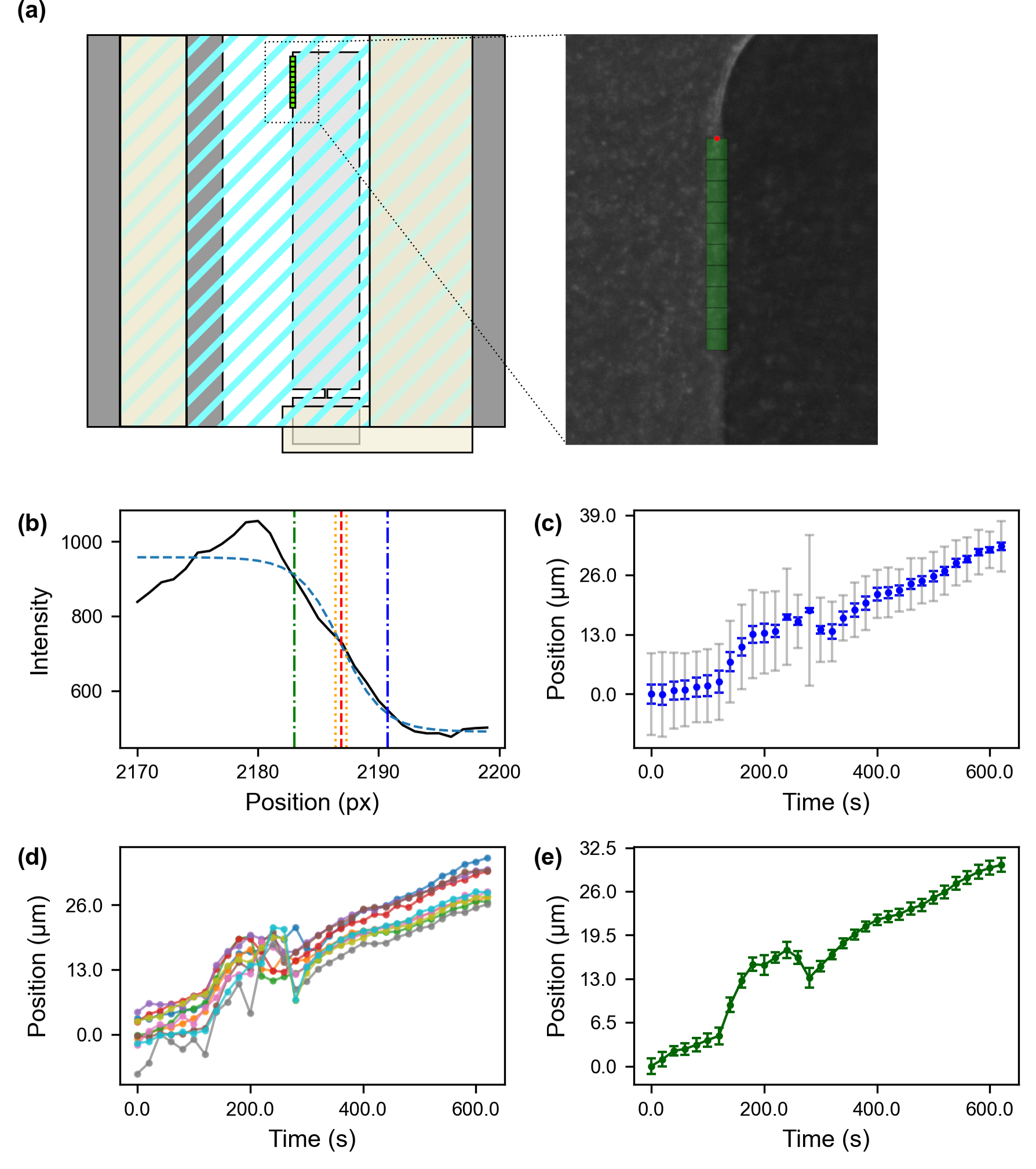}
    \caption{Flexure Edge Tracking \\ Flexure position is tracked through a statistical analysis of the intensity profiles along its edge. \\ (a) Ten sequential regions of interest (ROI, green squares) taken from the upper vertical edge of the flexure in the experimental chamber (cf. Fig \ref{fig:flexure}). Zoomed in data representation (dashed lines) \\ (b) Horizontal line scan of the average intensities for each pixel column of one ROI at a single representative frame. Black line: raw data; dashed blue line: fitted logistic curve; dashed red line: estimated edge position (inflection point of the logistic curve); dotted orange line: standard error of inflection point; green dot-dashed line: 90\% threshold; blue dot-dashed line:10\% threshold; 10\% to 90\% threshold defining "edge sharpness." \\ (c) Estimated flexure position relative to initial position in microns with standard error (blue) and edge sharpness (gray) across all frames for a single representative ROI. \\ (d) Estimated flexure position relative to the mean initial position in microns across all frames for all ten ROIs. \\ (e) Mean estimated flexure position at each frame for all ROIs  with standard error}
    \label{fig:position}
\end{figure}

\subsection{Edge tracking}
With these known values for the model experiment, we need the displacement of the flexure for force calculations, which we determine by tracking the edge of the flexure.
We track the the edge from the initial frame to until the gel is not clearly bound on the surface of the fleuxre or the opposite wall, which can happen as the gel contracts and slides off the boundary surfaces or rips off from the boundary surfaces.

Using ImageJ, all images are rotated with reference to the initial frame so that the lateral edge of the flexure is vertical. 
Then, an anchor point at the top of the flexure is selected, to which we position ten regions of interests. 
Each region of interest is a square of 30 x 30 pixels, with the top border of the first ROI centered at the anchor point and the following ROIs in a vertical column underneath it, as seen in Fig. \ref{fig:position}a. 

For each ROI, the intensities of all pixels in a vertical column are averaged, so that $\bar{I}=\frac{\sum_{i}^{n}I_i}{n}$ where $\bar{I}$ is the average intensity of a column of pixels, $I_i$ is the intensity of $i$-th pixel of a given column of pixels, and $n$ is the number of pixels in the column. 
Fig. \ref{fig:position}b shows $\bar{I}$ plotted across the horizontal X position for one frame.
A logistic curve $y = \frac{L}{1+e^{-k(x-x_0)}}+b$ is fit onto the raw data, and the predicted edge position is $x_0$. 
Standard error is drawn from the covariance matrix, and the 10\% to 90\% threshold define edge sharpness, or the length of the transition of the edge between the flexure and the sample chamber due to the blurred edges of the actomyosin gel overlapping with the flexure.
This is repeated for each frame in the experiment (Fig. \ref{fig:position}c), and repeated for each ROI (Fig. \ref{fig:position}d). 
The mean edge position relative to its initial position can be seen in Fig. \ref{fig:position}e. 

\begin{figure}
    \centering
    \includegraphics[width=0.5\linewidth]{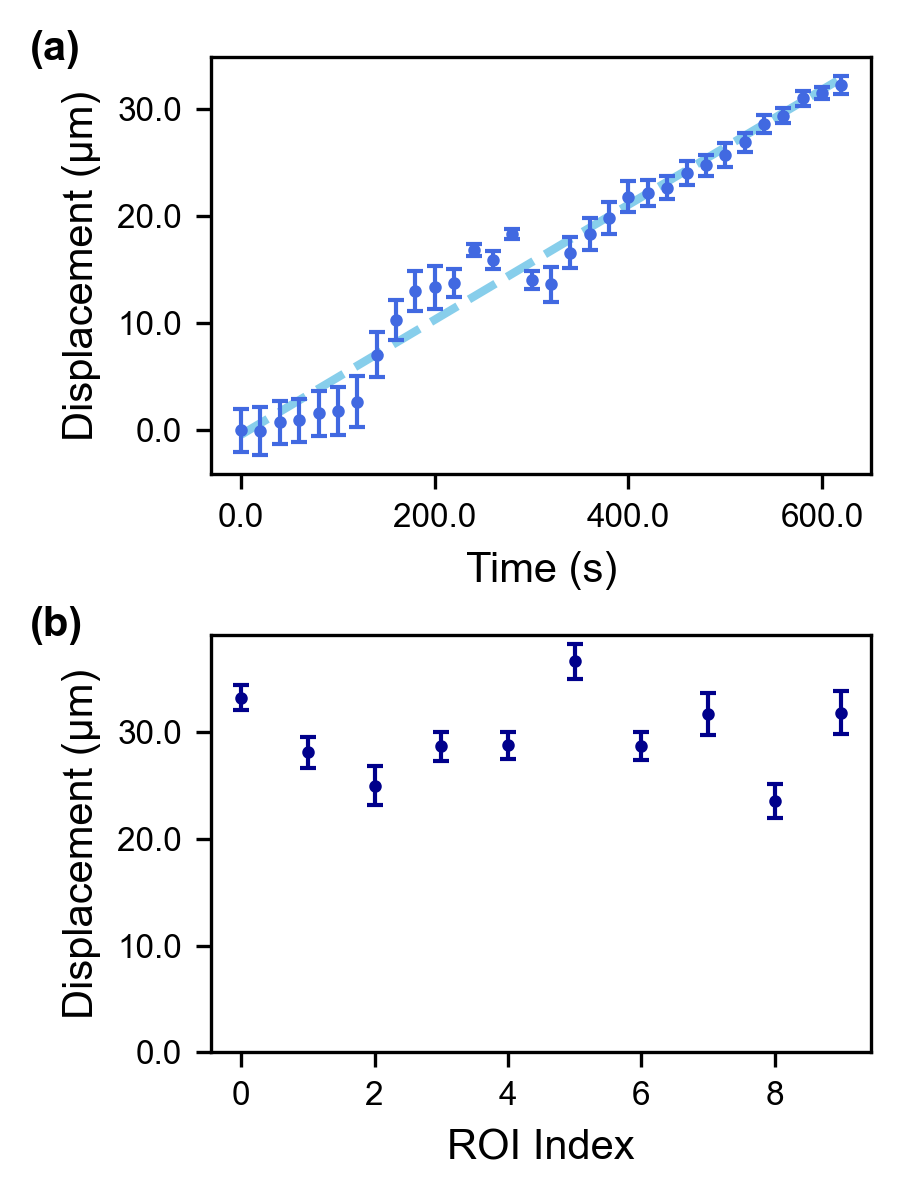}
    \caption{Flexure Displacement \\ Displacement is calculated using a linear model fit to the edge position estimations of each ROI. \\ (a) Linear fit (dashed line) of the estimated flexure position relative to initial position of a single ROI across all frames. \\ (b) Calculated displacement of one flexure from each ROI based on linear fits.}
    \label{fig:displacement}
\end{figure}

With the mean edge positions of the flexure, we calculate the displacement by performing a linear fit (Fig. \ref{fig:displacement}a), establishing the initial and final edge positions of each ROI and calculating the displacement as the difference between them (Fig. \ref{fig:displacement}b). 

The model experiment calculated an estimated displacement of \qty{29.665}{\micro\meter} $\pm$ \qty{0.468}{\micro\meter} of the flexure hinge. 

\subsection{Force calculation}
Force can be calculated by applying Hooke's law for the flexure, where \[F_{\rm{flexure}}=k_{\rm{flexure}} * \Delta x\] yielding \qty{11.962}{\micro\newton} $\pm$ \qty{0.495}{\micro\newton}.

Work done on the flexure by the gel $W$ can be calculated using the expression
\[W = \frac{1}{2}k_{\rm{flexure}}x^2 \]
where $\kflex$ is the spring constant of the flexure and $x$ is the displacement of the flexure, which yields $\qty{1778}{\pico\joule} \pm \qty{5.61}{\pico\joule}$.

\section{Discussion}

Here we have demonstrated methods for control over surface anchoring of reconstituted actomyosin networks.
We adapted an existing anchoring technique, based on oxidation of surface-bound PEG and attachment of actin-nucleating proteins \cite{reymann_nucleation_2010}, to two opposing surfaces.
This arrangement resulted in transverse anchoring, which differs from the lateral anchoring that has been more commonly studied.
Whereas lateral anchoring mimics actomyosin adhesion to the plasma membrane, transverse anchoring more closely mimics a supracellular contractile network coupling to deformable tissue.
Further, we applied our boundary-adhesion control methods to rigid boundaries (pinned contraction) as well as to one rigid and one flexible boundary (coupled contraction).

\subsection{Pinned Experiment}

Pinned contraction presents an opportunity to study the reciprocal mechanical interactions between stress-generating actomyosin active gels and a rigid mechanical environment.
The transverse anchoring we have established enables the generation of uniaxial, directional stresses in actin networks that would otherwise lack intrinsic directionality during free contraction.
In this light, transverse anchoring of contractile active gels more closely resembles tensile testing in conventional materials as opposed to traction force microscopy.
Further, contrasting transverse anchoring with free contraction offers an opportunity to systematically study anisotropy and symmetry-breaking mechanisms when actomyosin contractility interacts with its mechanical environment \cite{amiri_intracellular_2023, vogel_symmetry_2020, naganathan_active_2014, ierushalmi_centering_2020}.

Studying pinned contraction could further permit investigation of actomyosin gels under extreme stresses and strains, approaching the threshold of failure.
Understanding mechanical failure will help reveal how cytoskeletal networks withstand mechanical stress and avoid failure in living systems.
This behavior is relevant to understanding cellular resilience, tissue integrity, and morphogenesis \cite{koenderink_active_2009, brugues_forces_2014, pandya_actomyosin_2017, vedula_mechanics_2015, valencia_force-dependent_2021, valencia_actin_2025, sala_stress_2021, smith_zyxin-mediated_2010}.
Furthermore, fractures in actin networks not only help reveal determinants of cytoskeletal strength and failure thresholds, but have emerged as critical mechanosensitive elements that can facilitate further contraction by relieving built-up stress \cite{zsolnay_cracked_2024, matsuda_myosininduced_2024}.
Such fractures have been predominantly characterized under external force application; however, the current approach uniquely allows observation of fracture initiation and progression driven solely by internal myosin motor activity \cite{duque_rupture_2024}.

\subsection{Flexure Experiment}
The flexure chamber developed here represents a methodological advance for quantifying contractility in reconstituted actomyosin systems.
By calibrating the stiffness of the flexure, we converted a measured deflection of the flexure to a measure of the force exerted by the gel on the flexure, as well as the mechanical work output.
Our method will advance a more quantitative understanding of actomyosin active gels in their natural, biological context of actuation.

We measured a contraction force in the micronewton range (\qty{12.0}{\micro\newton} $\pm$ \qty{0.5}{\micro\newton}) which results in a net mechanical work of (\qty{1778}{\pico\joule} $\pm$ \qty{6}{\pico\joule}).
The force we measure is consistent with prior studies.
Early work, which anchored actomyosin active gels transversely to oil droplets, measured forces on the order of \qty{1}{\micro\newton} \cite{bendix_quantitative_2008}.
Force was estimated by applying the Young-Laplace equation to the oil-water interface and measuring the radius of curvature.
Another study investigated bundles of actomyosin anchored transversely between pairs of beads anchored to a polyacrylamide gel with known stiffness \cite{thoresen_reconstitution_2011}.
This resulted in measurements of the stall force of bundles of \qty{1}{\nano\newton}.
Meanwhile, actomyosin stresses were measured in a shear rheometer \cite{koenderink_active_2009}.
Further, mechanical power was measured in actomyosin networks in vesicles that either freely contracted \cite{sakamoto_f-actin_2024}, or were coupled to the inner surface of the vesicle \cite{sakamoto_mechanical_2024}.
In the free contraction case, power was measured on the order of \qty{0.01}{\pico\watt}, while in the coupled case it increased to \qty{10}{\pico\watt}.
In these studies, deformation was measured, and force was determined based on scaling arguments.
This combination resulted in power measurements that are consistent with prior studies.

Flexure hinges have the potential to offer several advantages as force sensors for future studies of contractile actomyosin.
The force range of \qtyrange{1}{10}{\micro\newton} that actomyosin gels exert are too large for most AFM cantilevers, but are too small for the most sensitive load cells.
Flexure hinges are easily combined with fluorescence microscopy techniques, allowing one to measure force and actomyosin gel configuration simultaneously.
Further, the sample can be activated with light, allowing for optogenetic manipulation.
The bending response of flexure hinges rests crucially on the design of the neck region.
Control over the geometry allows one to systematically tune the stiffness of the cantilever.
Furthermore, flexure hinges can be easily designed to exhibit a constant stiffness over a broader range of deformations, compared to cantilevered beams of uniform cross section.
These advantages make flexure hinges a suitable candidate toward precise measurements of force in contexts where actomyosin gels exhibit large strains.
Precise force measurements would enable a more complete characterization of the active force-generating properties of actomyosin active gels.
Despite these advantages, flexure hinges require additional optimization in order to provide high sensitivity.
One notable path for technical optimization is the incorporation of high-resolution brightfield microscopy for edge tracking.
This would allow more precise segmentation and analysis of the gel–flexure interface, improving the accuracy of edge tracking.
An additional improvement would be to explore alternative materials for the flexure hinge.
In our experience, poly methyl methacrylate (PMMA) hinges are brittle and prone to failure during sample preparation.
Materials with a higher strength and yield stress, such as metals, could offer improvements.

Flexure hinges offer an alternative to traction force microscopy, which is a well-established method for measuring force.
TFM relies on lateral anchoring, where adhesions points are distributed over one broad, planar interface.
TFM is a powerful technique for cell culture applications, such as mapping the tractional “footprint” of a migrating cell or the forces exerted by focal adhesions on elastic gels. 
Meanwhile, flexure hinges anchor gels transversely, which allows isotropic stresses inside the gel to rearrange, giving rise to uniaxial stresses.
Furthermore, flexure hinges permit large strains.
Large strains are advantageous, because mechanical work (or power) are only achieved at finite stresses and strains (or strain rates).
Gels strongly anchored laterally to a gel undergoing little strain would not exert significant work.
In this scenario, motors would likely operate near stall conditions, where strain rate and thus power are zero.
Flexure hinges undergoing large strains offer a platform for accurate measurements not only of force, but also mechanical work and power.

\subsection{Outlook}
The pairing of pinned and flexure experiments unlocks rich and complementary perspectives into the emergent mechanics of the actomyosin network. 
The rigid boundary system excels at illuminating fundamental mechanisms: how forces organize and build up within a network fixed at the borders, how anisotropic contraction gives rise to stress distribution and internal rearrangement, and potentially how filaments respond to various nucleators, crosslinkers, or myosin isoforms. 
This configuration mimics settings of strong tissue confinement found in wound edges or epithelial sheets, where mechanical feedback and rupture may dominate the response to internal contraction.

In contrast, the flexure assay specializes in resolving the total contractile force output and energetic efficiency of the network against a compliant boundary condition, emulating scenarios such as the deformation of soft connective tissues. 
This flexure system is modular, allowing easy adjustment, customization, and extension of its components to suit different experimental needs or scenarios.
By tuning the flexure’s stiffness, experimentalists can simulate environments ranging from soft, developmentally dynamic tissues to stiffer, more mature extracellular matrices, gaining insight into how cellular mechanics adapt to external loads. 
In addition, the composition of the actomyosin assay can be altered to investigate the relationship between specific protein concentration and force generation. 
The experimental chamber can be resized or reconfigured to accommodate different gel sizes, actomyosin architectures, or even multicellular aggregates, depending on research goals.
Multiple flexure units can be deployed in parallel, enabling higher-throughput assays, comparative studies, or multiplexed perturbations within the same experimental framework.
This flexibility makes the flexure system especially powerful for systematic, comparative, or multi-condition experimentation—unlike rigid, single-purpose devices, a modular system can evolve alongside scientific questions and technological innovations.

Both platforms together point the way toward more physiologically relevant models of cellular biomechanics. The simplicity and accuracy of the flexure method—combined with the spatial control and visualization afforded by pinned boundaries—support the investigation of new phenomena: load-dependent cytoskeletal remodeling, fracture mechanics, and self-organization under complex boundary geometries.

Moving forward, both strategies could be further enhanced by integration of high-speed, high-resolution imaging  for collective analysis of morphological, compositional, and force signatures.
Coupling mechanical measurements with biochemical markers like zyxin \cite{Sun-2020-DevCell}, a mechanosensitive protein that responds to tension in actin filaments could also yield further insight into stress distribution and internal rearrangement.
Stronger and more consistent anchoring protocols, perhaps using biotin, that establish stronger adhesion between the actin network and the boundary conditions will also provide more accurate and reproducible data. 
Adapting the setup for in vivo–like environments, such as 3D matrices or co-culture systems to bridge the gap from reconstitution to tissue mechanics.

\section{Conclusion}

Here we have reported methods that allow actomyosin active gels to adhere to boundaries in a controlled fashion. We have developed methods to selectively anchor contractile actomyosin to two opposing, rigid surfaces, as well as to one rigid and one flexible surface. The flexible surface was implemented by means of a flexure hinge, a deformable structure with tunable geometry and thus stiffness. Contraction against a flexible surface allows contractile active gels to exert mechanical work. We were able to provide an initial estimate of \qty{1778}{\pico\joule} of work done by a gel on a flexure. We anticipate that future refinements to this method will allow for improved characterization of actomyosin mechanical properties and actuation response.

\section{Acknowledgements}

This work was primarily supported by the National Science Foundation under DMR-2308817. We acknowledge further partial support under National Science Foundation through the Center for Dynamics and Control of Materials: an NSF MRSEC under Cooperative Agreement No. DMR-1720595; as well as the National Institutes of Health under NICHD R21HD112657. We thank Jeff Boney for manufacture of the flexure hinges, and Omar Cantu for advice on developing the surface adhesion protocol.

\printbibliography

@article{Lieleg-2010-SoftMatter,
  title = {Structure and Dynamics of Cross-Linked Actin Networks},
  author = {Lieleg, Oliver and Claessens, Mireille M A E and Bausch, Andreas R},
  year = 2010,
  journal = {Soft Matter},
  volume = {6},
  number = {2},
  pages = {218 225},
  issn = {1744-683X},
  doi = {10.1039/b912163n},
  langid = {english}
}

@article{torii1996,
  title = {A Method for Determining the Spring Constant of Cantilevers for Atomic Force Microscopy},
  author = {Torii, Akihiro and Sasaki, Minoru and Hane, Kazuhiro and Okuma, Shigeru},
  year = 1996,
  month = feb,
  journal = {Meas. Sci. Technol.},
  volume = {7},
  number = {2},
  pages = {179--184},
  publisher = {IOP Publishing},
  issn = {0957-0233},
  doi = {10.1088/0957-0233/7/2/010},
  urldate = {2022-03-23},
  langid = {english},
 }

@article{Sader99,
  ids = {sader1999},
  title = {Calibration of Rectangular Atomic Force Microscope Cantilevers},
  author = {Sader, John E. and Chon, James W.M. and Mulvaney, Paul},
  year = 1999,
  journal = {Rev. Sci. Instrum.},
  volume = {70},
  number = {10},
  pages = {3967--3969},
  issn = {00346748},
  doi = {10.1063/1.1150021}
}

@article{Sun-2020-DevCell,
  title = {Mechanosensing through {{Direct Binding}} of {{Tensed F-Actin}} by {{LIM Domains}}},
  author = {Sun, Xiaoyu and Phua, Donovan Y. Z. and Axiotakis, Lucas and Smith, Mark A. and Blankman, Elizabeth and Gong, Rui and Cail, Robert C. and {Espinosa de Los Reyes}, Santiago and Beckerle, Mary C. and Waterman, Clare M. and Alushin, Gregory M.},
  year = 2020,
  month = nov,
  journal = {Developmental Cell},
  volume = {55},
  number = {4},
  pages = {468-482.e7},
  issn = {1878-1551},
  doi = {10.1016/j.devcel.2020.09.022},
  langid = {english},
  pmcid = {PMC7686152},
  pmid = {33058779}
}

@article{Movilli-2020-Org.Mater.,
  title = {Functionalized {{Polyelectrolytes}} for {{Bioengineered Interfaces}} and {{Biosensing Applications}}},
  author = {Movilli, Jacopo and Huskens, Jurriaan},
  year = 2020,
  month = apr,
  journal = {Organic Materials},
  volume = {02},
  pages = {78--107},
  publisher = {Georg Thieme Verlag KG},
  doi = {10.1055/s-0040-1708494},
  urldate = {2024-09-24},
  copyright = {Georg Thieme Verlag KG R\"udigerstra\ss e 14, 70469 Stuttgart, Germany}
}

@article{koenderink_active_2009,
	title = {An active biopolymer network controlled by molecular motors},
	volume = {106},
	issn = {0027-8424},
	url = {http://www.pnas.org/content/106/36/15192},
	doi = {10.1073/pnas.0903974106},
	language = {English},
	number = {36},
	journal = {Proceedings Of The National Academy Of Sciences Of The United States Of America},
	author = {Koenderink, Gijsje H and Dogic, Zvonimir and Nakamura, Fumihiko and Bendix, Poul M and MacKintosh, Frederick C and Hartwig, John H and Stossel, Thomas P and Weitz, David A},
	year = {2009},
	pages = {15192 15197},
}

@article{thoresen_reconstitution_2011,
	title = {Reconstitution of {Contractile} {Actomyosin} {Bundles}},
	volume = {100},
	issn = {0006-3495},
	doi = {10.1016/j.bpj.2011.04.031},
	number = {11},
	journal = {Biophysical Journal},
	author = {Thoresen, Todd and Lenz, Martin and Gardel, Margaret L.},
	year = {2011},
	pages = {2698--2705},
}

@article{reymann_actin_2012,
	title = {Actin network architecture can determine myosin motor activity.},
	volume = {336},
	issn = {0036-8075},
	url = {https://app.readcube.com/library/42a883be-d4df-4536-8337-8723b4559f27/item/81B9615B-8512-44DD-858A-1372D2C9C57F},
	doi = {10.1126/science.1221708},
	language = {English},
	number = {6086},
	journal = {Science},
	author = {Reymann, Anne-Cécile and Boujemaa-Paterski, Rajaa and Martiel, Jean-Louis and Guerin, Christophe and Cao, Wenxiang and Chin, Harvey F and Cruz, Enrique M De La and Théry, Manuel and Blanchoin, Laurent},
	month = jun,
	year = {2012},
	pages = {1310 1314},
}

@article{murrell_f-actin_2012,
	title = {F-actin buckling coordinates contractility and severing in a biomimetic actomyosin cortex},
	volume = {109},
	issn = {0027-8424},
	url = {https://app.readcube.com/library/42a883be-d4df-4536-8337-8723b4559f27/item/96E7AE28-6CB2-4628-8210-61AC3965C932},
	doi = {10.1073/pnas.1214753109},
	language = {English},
	number = {51},
	journal = {Proceedings Of The National Academy Of Sciences Of The United States Of America},
	author = {Murrell, Michael P and Gardel, Margaret L},
	month = jan,
	year = {2012},
	pages = {20820 20825},
}

@article{reymann_nucleation_2010,
	title = {Nucleation geometry governs ordered actin networks structures},
	volume = {9},
	issn = {1476-4660},
	doi = {10.1038/nmat2855},
	number = {10},
	journal = {Nature Materials},
	author = {Reymann, Anne-Cécile and Martiel, Jean-Louis and Cambier, Théo and Blanchoin, Laurent and Boujemaa-Paterski, Rajaa and Théry, Manuel},
	month = sep,
	year = {2010},
	pages = {827 832},
}

@book{lobontiu_compliant_2020,
	address = {Boca Raton},
	edition = {2},
	title = {Compliant {Mechanisms}: {Design} of {Flexure} {Hinges}},
	isbn = {978-0-429-18466-6},
	shorttitle = {Compliant {Mechanisms}},
	publisher = {CRC Press},
	author = {Lobontiu, Nicolae},
	month = nov,
	year = {2020},
	doi = {10.1201/9780429184666},
}

@article{bendix_quantitative_2008,
	title = {A quantitative analysis of contractility in active cytoskeletal protein networks.},
	volume = {94},
	issn = {0006-3495},
	doi = {10.1529/biophysj.107.117960},
	language = {English},
	number = {8},
	journal = {Biophysical Journal},
	author = {Bendix, Poul M and Koenderink, Gijsje H and Cuvelier, Damien and Dogic, Zvonimir and Koeleman, Bernard N and Brieher, William M and Field, Christine M and Mahadevan, L and Weitz, David A},
	month = apr,
	year = {2008},
	pages = {3126 3136},
}

@article{paluch_role_2013,
	title = {The role and regulation of blebs in cell migration},
	volume = {25},
	issn = {09550674},
	url = {https://linkinghub.elsevier.com/retrieve/pii/S0955067413000793},
	doi = {10.1016/j.ceb.2013.05.005},
	language = {en},
	number = {5},
	urldate = {2025-10-29},
	journal = {Current Opinion in Cell Biology},
	author = {Paluch, Ewa K and Raz, Erez},
	month = oct,
	year = {2013},
	pages = {582--590},
}

@article{pollard_cellular_2003,
	title = {Cellular {Motility} {Driven} by {Assembly} and {Disassembly} of {Actin} {Filaments}},
	volume = {112},
	issn = {00928674},
	url = {https://linkinghub.elsevier.com/retrieve/pii/S009286740300120X},
	doi = {10.1016/S0092-8674(03)00120-X},
	language = {en},
	number = {4},
	urldate = {2025-10-29},
	journal = {Cell},
	author = {Pollard, Thomas D and Borisy, Gary G},
	month = feb,
	year = {2003},
	pages = {453--465},
}

@article{murrell_forcing_2015,
	title = {Forcing cells into shape: the mechanics of actomyosin contractility},
	volume = {16},
	issn = {1471-0072, 1471-0080},
	shorttitle = {Forcing cells into shape},
	url = {https://www.nature.com/articles/nrm4012},
	doi = {10.1038/nrm4012},
	language = {en},
	number = {8},
	urldate = {2025-10-29},
	journal = {Nature Reviews Molecular Cell Biology},
	author = {Murrell, Michael and Oakes, Patrick W. and Lenz, Martin and Gardel, Margaret L.},
	month = aug,
	year = {2015},
	pages = {486--498},
}

@article{pandya_actomyosin_2017,
	title = {Actomyosin contractility and collective migration: may the force be with you},
	volume = {48},
	issn = {09550674},
	shorttitle = {Actomyosin contractility and collective migration},
	url = {https://linkinghub.elsevier.com/retrieve/pii/S0955067417300261},
	doi = {10.1016/j.ceb.2017.06.006},
	language = {en},
	urldate = {2025-10-29},
	journal = {Current Opinion in Cell Biology},
	author = {Pandya, Pahini and Orgaz, Jose L and Sanz-Moreno, Victoria},
	month = oct,
	year = {2017},
	pages = {87--96},
}

@article{brugues_forces_2014,
	title = {Forces driving epithelial wound healing},
	volume = {10},
	issn = {1745-2473, 1745-2481},
	url = {https://www.nature.com/articles/nphys3040},
	doi = {10.1038/nphys3040},
	language = {en},
	number = {9},
	urldate = {2025-10-29},
	journal = {Nature Physics},
	author = {Brugués, Agustí and Anon, Ester and Conte, Vito and Veldhuis, Jim H. and Gupta, Mukund and Colombelli, Julien and Muñoz, José J. and Brodland, G. Wayne and Ladoux, Benoit and Trepat, Xavier},
	month = sep,
	year = {2014},
	pages = {683--690},
}

@article{vedula_mechanics_2015,
	title = {Mechanics of epithelial closure over non-adherent environments},
	volume = {6},
	issn = {2041-1723},
	url = {https://www.nature.com/articles/ncomms7111},
	doi = {10.1038/ncomms7111},
	language = {en},
	number = {1},
	urldate = {2025-10-29},
	journal = {Nature Communications},
	author = {Vedula, Sri Ram Krishna and Peyret, Grégoire and Cheddadi, Ibrahim and Chen, Tianchi and Brugués, Agustí and Hirata, Hiroaki and Lopez-Menendez, Horacio and Toyama, Yusuke and Neves De Almeida, Luís and Trepat, Xavier and Lim, Chwee Teck and Ladoux, Benoit},
	month = jan,
	year = {2015},
	pages = {6111},
}

@article{stachowiak_mechanism_2014,
	title = {Mechanism of {Cytokinetic} {Contractile} {Ring} {Constriction} in {Fission} {Yeast}},
	volume = {29},
	issn = {15345807},
	url = {https://linkinghub.elsevier.com/retrieve/pii/S153458071400241X},
	doi = {10.1016/j.devcel.2014.04.021},
	language = {en},
	number = {5},
	urldate = {2025-10-29},
	journal = {Developmental Cell},
	author = {Stachowiak, Matthew R. and Laplante, Caroline and Chin, Harvey F. and Guirao, Boris and Karatekin, Erdem and Pollard, Thomas D. and O’Shaughnessy, Ben},
	month = jun,
	year = {2014},
	pages = {547--561},
}

@article{svitkina_actin_2018,
	title = {The {Actin} {Cytoskeleton} and {Actin}-{Based} {Motility}},
	volume = {10},
	issn = {1943-0264},
	url = {http://cshperspectives.cshlp.org/lookup/doi/10.1101/cshperspect.a018267},
	doi = {10.1101/cshperspect.a018267},
	language = {en},
	number = {1},
	urldate = {2025-10-29},
	journal = {Cold Spring Harbor Perspectives in Biology},
	author = {Svitkina, Tatyana},
	month = jan,
	year = {2018},
	pages = {a018267},
}

@article{kasza_dynamics_2011,
	title = {Dynamics and regulation of contractile actin–myosin networks in morphogenesis},
	volume = {23},
	issn = {09550674},
	url = {https://linkinghub.elsevier.com/retrieve/pii/S0955067410001857},
	doi = {10.1016/j.ceb.2010.10.014},
	language = {en},
	number = {1},
	urldate = {2025-10-29},
	journal = {Current Opinion in Cell Biology},
	author = {Kasza, Karen E and Zallen, Jennifer A},
	month = feb,
	year = {2011},
	pages = {30--38},
}

@article{style_traction_2014,
	title = {Traction force microscopy in physics and biology},
	volume = {10},
	issn = {1744-6848},
	url = {https://pubs.rsc.org/en/content/articlelanding/2014/sm/c4sm00264d},
	doi = {10.1039/C4SM00264D},
	language = {en},
	number = {23},
	urldate = {2025-01-20},
	journal = {Soft Matter},
	author = {Style, Robert W. and Boltyanskiy, Rostislav and German, Guy K. and Hyland, Callen and MacMinn, Christopher W. and Mertz, Aaron F. and Wilen, Larry A. and Xu, Ye and Dufresne, Eric R.},
	month = may,
	year = {2014},
	note = {Publisher: The Royal Society of Chemistry},
	pages = {4047--4055},
}

@article{duque_rupture_2024,
	title = {Rupture strength of living cell monolayers},
	volume = {23},
	issn = {1476-1122, 1476-4660},
	url = {https://www.nature.com/articles/s41563-024-02027-3},
	doi = {10.1038/s41563-024-02027-3},
	language = {en},
	number = {11},
	urldate = {2025-10-29},
	journal = {Nature Materials},
	author = {Duque, Julia and Bonfanti, Alessandra and Fouchard, Jonathan and Baldauf, Lucia and Azenha, Sara R. and Ferber, Emma and Harris, Andrew and Barriga, Elias H. and Kabla, Alexandre J. and Charras, Guillaume},
	month = nov,
	year = {2024},
	pages = {1563--1574},
}

@article{matsuda_myosininduced_2024,
	title = {Myosin‐induced {F}‐actin fragmentation facilitates contraction of actin networks},
	volume = {81},
	issn = {1949-3584, 1949-3592},
	url = {https://onlinelibrary.wiley.com/doi/10.1002/cm.21848},
	doi = {10.1002/cm.21848},
	language = {en},
	number = {8},
	urldate = {2025-10-29},
	journal = {Cytoskeleton},
	author = {Matsuda, Kyohei and Jung, Wonyeong and Sato, Yusei and Kobayashi, Takuya and Yamagishi, Masahiko and Kim, Taeyoon and Yajima, Junichiro},
	month = aug,
	year = {2024},
	pages = {339--355},
}

@article{zsolnay_cracked_2024,
	title = {Cracked actin filaments as mechanosensitive receptors},
	volume = {123},
	issn = {00063495},
	url = {https://linkinghub.elsevier.com/retrieve/pii/S0006349524004119},
	doi = {10.1016/j.bpj.2024.06.014},
	language = {en},
	number = {19},
	urldate = {2025-10-29},
	journal = {Biophysical Journal},
	author = {Zsolnay, Vilmos and Gardel, Margaret L. and Kovar, David R. and Voth, Gregory A.},
	month = oct,
	year = {2024},
	pages = {3283--3294},
}

@incollection{alvarado_reconstituting_2015,
	title = {Reconstituting cytoskeletal contraction events with biomimetic actin–myosin active gels},
	volume = {128},
	copyright = {https://www.elsevier.com/tdm/userlicense/1.0/},
	isbn = {978-0-12-802450-8},
	url = {https://linkinghub.elsevier.com/retrieve/pii/S0091679X15000643},
	language = {en},
	urldate = {2025-11-05},
	booktitle = {Methods in {Cell} {Biology}},
	publisher = {Elsevier},
	author = {Alvarado, José and Koenderink, Gijsje H.},
	year = {2015},
	doi = {10.1016/bs.mcb.2015.02.001},
	pages = {83--103},
}

@article{schmidt_safe_2022,
	title = {Safe {Piranhas}: {A} {Review} of {Methods} and {Protocols}},
	volume = {29},
	copyright = {https://doi.org/10.15223/policy-029},
	issn = {1871-5532, 1878-0504},
	shorttitle = {Safe {Piranhas}},
	url = {https://pubs.acs.org/doi/10.1021/acs.chas.1c00094},
	doi = {10.1021/acs.chas.1c00094},
	language = {en},
	number = {1},
	urldate = {2025-11-05},
	journal = {ACS Chemical Health \& Safety},
	author = {Schmidt, Hugo Gerald},
	month = jan,
	year = {2022},
	pages = {54--61},
}

@article{arima_complement_2008,
	title = {Complement activation on surfaces modified with ethylene glycol units},
	volume = {29},
	issn = {01429612},
	url = {https://linkinghub.elsevier.com/retrieve/pii/S0142961207008204},
	doi = {10.1016/j.biomaterials.2007.10.015},
	language = {en},
	number = {5},
	urldate = {2025-11-05},
	journal = {Biomaterials},
	author = {Arima, Yusuke and Toda, Mitsuaki and Iwata, Hiroo},
	month = feb,
	year = {2008},
	pages = {551--560},
}

@article{blanchoin_actin_2014,
	title = {Actin {Dynamics}, {Architecture}, and {Mechanics} in {Cell} {Motility}},
	volume = {94},
	issn = {0031-9333, 1522-1210},
	url = {https://www.physiology.org/doi/10.1152/physrev.00018.2013},
	doi = {10.1152/physrev.00018.2013},
	language = {en},
	number = {1},
	urldate = {2025-11-05},
	journal = {Physiological Reviews},
	author = {Blanchoin, Laurent and Boujemaa-Paterski, Rajaa and Sykes, Cécile and Plastino, Julie},
	month = jan,
	year = {2014},
	pages = {235--263},
}

@article{soares_e_silva_active_2011,
	title = {Active multistage coarsening of actin networks driven by myosin motors},
	volume = {108},
	issn = {0027-8424, 1091-6490},
	url = {https://pnas.org/doi/full/10.1073/pnas.1016616108},
	doi = {10.1073/pnas.1016616108},
	language = {en},
	number = {23},
	urldate = {2025-11-05},
	journal = {Proceedings of the National Academy of Sciences},
	author = {Soares E Silva, Marina and Depken, Martin and Stuhrmann, Björn and Korsten, Marijn and MacKintosh, Fred C. and Koenderink, Gijsje H.},
	month = jun,
	year = {2011},
	pages = {9408--9413},
}

@article{paluch_dynamic_2006,
	title = {Dynamic modes of the cortical actomyosin gel during cell locomotion and division},
	volume = {16},
	copyright = {https://www.elsevier.com/tdm/userlicense/1.0/},
	issn = {09628924},
	url = {https://linkinghub.elsevier.com/retrieve/pii/S0962892405002965},
	doi = {10.1016/j.tcb.2005.11.003},
	language = {en},
	number = {1},
	urldate = {2025-11-23},
	journal = {Trends in Cell Biology},
	author = {Paluch, Ewa and Sykes, Cécile and Prost, Jacques and Bornens, Michel},
	month = jan,
	year = {2006},
	pages = {5--10},
}

@article{koenderink_architecture_2018,
	title = {Architecture shapes contractility in actomyosin networks},
	volume = {50},
	issn = {09550674},
	url = {https://linkinghub.elsevier.com/retrieve/pii/S0955067418300061},
	doi = {10.1016/j.ceb.2018.01.015},
	language = {en},
	urldate = {2025-11-23},
	journal = {Current Opinion in Cell Biology},
	author = {Koenderink, Gijsje H and Paluch, Ewa K},
	month = feb,
	year = {2018},
	pages = {79--85},
}

@incollection{hinz_quantitative_2021,
	address = {New York, NY},
	title = {Quantitative {Analysis} of {Myofibroblast} {Contraction} by {Traction} {Force} {Microscopy}},
	volume = {2299},
	isbn = {978-1-0716-1381-8 978-1-0716-1382-5},
	url = {https://link.springer.com/10.1007/978-1-0716-1382-5_14},
	language = {en},
	urldate = {2025-11-23},
	booktitle = {Myofibroblasts},
	publisher = {Springer US},
	author = {Yang, Shuying and Valencia, Fernando R. and Sabass, Benedikt and Plotnikov, Sergey V.},
	editor = {Hinz, Boris and Lagares, David},
	year = {2021},
	doi = {10.1007/978-1-0716-1382-5_14},
	note = {Series Title: Methods in Molecular Biology},
	pages = {181--195},
}

@article{de_la_pena_quantifying_2021,
	title = {Quantifying cellular forces: {Practical} considerations of traction force microscopy for dermal fibroblasts},
	volume = {30},
	issn = {0906-6705, 1600-0625},
	shorttitle = {Quantifying cellular forces},
	url = {https://onlinelibrary.wiley.com/doi/10.1111/exd.14166},
	doi = {10.1111/exd.14166},
	language = {en},
	number = {1},
	urldate = {2025-11-23},
	journal = {Experimental Dermatology},
	author = {De La Pena, Abigail and Mukhtar, Marah and Yokosawa, Ryosuke and Carrasquilla, Santiago and Simmons, Chelsey S.},
	month = jan,
	year = {2021},
	pages = {74--83},
}

@article{lo_cell_2000,
	title = {Cell {Movement} {Is} {Guided} by the {Rigidity} of the {Substrate}},
	volume = {79},
	copyright = {https://www.elsevier.com/tdm/userlicense/1.0/},
	issn = {00063495},
	url = {https://linkinghub.elsevier.com/retrieve/pii/S0006349500762795},
	doi = {10.1016/S0006-3495(00)76279-5},
	language = {en},
	number = {1},
	urldate = {2025-11-23},
	journal = {Biophysical Journal},
	author = {Lo, Chun-Min and Wang, Hong-Bei and Dembo, Micah and Wang, Yu-li},
	month = jul,
	year = {2000},
	pages = {144--152},
}

@article{engler_matrix_2006,
	title = {Matrix {Elasticity} {Directs} {Stem} {Cell} {Lineage} {Specification}},
	volume = {126},
	copyright = {https://www.elsevier.com/tdm/userlicense/1.0/},
	issn = {00928674},
	url = {https://linkinghub.elsevier.com/retrieve/pii/S0092867406009615},
	doi = {10.1016/j.cell.2006.06.044},
	language = {en},
	number = {4},
	urldate = {2025-11-23},
	journal = {Cell},
	author = {Engler, Adam J. and Sen, Shamik and Sweeney, H. Lee and Discher, Dennis E.},
	month = aug,
	year = {2006},
	pages = {677--689},
}

@article{balaban_force_2001,
	title = {Force and focal adhesion assembly: a close relationship studied using elastic micropatterned substrates},
	volume = {3},
	copyright = {http://www.springer.com/tdm},
	issn = {1465-7392, 1476-4679},
	shorttitle = {Force and focal adhesion assembly},
	url = {https://www.nature.com/articles/ncb0501_466},
	doi = {10.1038/35074532},
	language = {en},
	number = {5},
	urldate = {2025-11-23},
	journal = {Nature Cell Biology},
	author = {Balaban, Nathalie Q. and Schwarz, Ulrich S. and Riveline, Daniel and Goichberg, Polina and Tzur, Gila and Sabanay, Ilana and Mahalu, Diana and Safran, Sam and Bershadsky, Alexander and Addadi, Lia and Geiger, Benjamin},
	month = may,
	year = {2001},
	pages = {466--472},
}

@article{trichet_evidence_2012,
	title = {Evidence of a large-scale mechanosensing mechanism for cellular adaptation to substrate stiffness},
	volume = {109},
	issn = {0027-8424, 1091-6490},
	url = {https://pnas.org/doi/full/10.1073/pnas.1117810109},
	doi = {10.1073/pnas.1117810109},
	language = {en},
	number = {18},
	urldate = {2025-11-23},
	journal = {Proceedings of the National Academy of Sciences},
	author = {Trichet, Léa and Le Digabel, Jimmy and Hawkins, Rhoda J. and Vedula, Sri Ram Krishna and Gupta, Mukund and Ribrault, Claire and Hersen, Pascal and Voituriez, Raphaël and Ladoux, Benoît},
	month = may,
	year = {2012},
	pages = {6933--6938},
}

@article{bangasser_determinants_2013,
	title = {Determinants of {Maximal} {Force} {Transmission} in a {Motor}-{Clutch} {Model} of {Cell} {Traction} in a {Compliant} {Microenvironment}},
	volume = {105},
	issn = {00063495},
	url = {https://linkinghub.elsevier.com/retrieve/pii/S0006349513007066},
	doi = {10.1016/j.bpj.2013.06.027},
	language = {en},
	number = {3},
	urldate = {2025-11-23},
	journal = {Biophysical Journal},
	author = {Bangasser, Benjamin L. and Rosenfeld, Steven S. and Odde, David J.},
	month = aug,
	year = {2013},
	pages = {581--592},
}

@article{panzetta_cell_2019,
	title = {Cell mechanosensing is regulated by substrate strain energy rather than stiffness},
	volume = {116},
	issn = {0027-8424, 1091-6490},
	url = {https://pnas.org/doi/full/10.1073/pnas.1904660116},
	doi = {10.1073/pnas.1904660116},
	language = {en},
	number = {44},
	urldate = {2025-11-23},
	journal = {Proceedings of the National Academy of Sciences},
	author = {Panzetta, Valeria and Fusco, Sabato and Netti, Paolo A.},
	month = oct,
	year = {2019},
	pages = {22004--22013},
}

@article{shi_pervasive_2021,
	title = {Pervasive cytoquakes in the actomyosin cortex across cell types and substrate stiffness},
	volume = {13},
	copyright = {https://academic.oup.com/journals/pages/open\_access/funder\_policies/chorus/standard\_publication\_model},
	issn = {1757-9708},
	url = {https://academic.oup.com/ib/article/13/10/246/6455026},
	doi = {10.1093/intbio/zyab017},
	language = {en},
	number = {10},
	urldate = {2025-11-23},
	journal = {Integrative Biology},
	author = {Shi, Yu and Sivarajan, Shankar and Xiang, Katherine M and Kostecki, Geran M and Tung, Leslie and Crocker, John C and Reich, Daniel H},
	month = dec,
	year = {2021},
	pages = {246--257},
}

@article{zhou_effects_2017,
	title = {Effects of substrate stiffness and actomyosin contractility on coupling between force transmission and vinculin–paxillin recruitment at single focal adhesions},
	volume = {28},
	issn = {1059-1524, 1939-4586},
	url = {https://www.molbiolcell.org/doi/10.1091/mbc.e17-02-0116},
	doi = {10.1091/mbc.e17-02-0116},
	language = {en},
	number = {14},
	urldate = {2025-11-23},
	journal = {Molecular Biology of the Cell},
	author = {Zhou, Dennis W. and Lee, Ted T. and Weng, Shinuo and Fu, Jianping and García, Andrés J.},
	editor = {Weaver, Valerie Marie},
	month = jul,
	year = {2017},
	pages = {1901--1911},
}

@article{dominguez_actin_2011,
	title = {Actin {Structure} and {Function}},
	volume = {40},
	issn = {1936-122X, 1936-1238},
	url = {https://www.annualreviews.org/doi/10.1146/annurev-biophys-042910-155359},
	doi = {10.1146/annurev-biophys-042910-155359},
	language = {en},
	number = {1},
	urldate = {2025-12-03},
	journal = {Annual Review of Biophysics},
	author = {Dominguez, Roberto and Holmes, Kenneth C.},
	month = jun,
	year = {2011},
	pages = {169--186},
}

@article{lappalainen_biochemical_2022,
	title = {Biochemical and mechanical regulation of actin dynamics},
	volume = {23},
	issn = {1471-0072, 1471-0080},
	url = {https://www.nature.com/articles/s41580-022-00508-4},
	doi = {10.1038/s41580-022-00508-4},
	language = {en},
	number = {12},
	urldate = {2025-12-03},
	journal = {Nature Reviews Molecular Cell Biology},
	author = {Lappalainen, Pekka and Kotila, Tommi and Jégou, Antoine and Romet-Lemonne, Guillaume},
	month = dec,
	year = {2022},
	pages = {836--852},
}

@article{wang_active_2012,
	title = {Active contractility in actomyosin networks},
	volume = {109},
	issn = {0027-8424, 1091-6490},
	url = {https://pnas.org/doi/full/10.1073/pnas.1204205109},
	doi = {10.1073/pnas.1204205109},
	language = {en},
	number = {17},
	urldate = {2025-12-03},
	journal = {Proceedings of the National Academy of Sciences},
	author = {Wang, Shenshen and Wolynes, Peter G.},
	month = apr,
	year = {2012},
	pages = {6446--6451},
}

@article{pollard_actin_2016,
	title = {Actin and {Actin}-{Binding} {Proteins}},
	volume = {8},
	issn = {1943-0264},
	url = {http://cshperspectives.cshlp.org/lookup/doi/10.1101/cshperspect.a018226},
	doi = {10.1101/cshperspect.a018226},
	language = {en},
	number = {8},
	urldate = {2025-12-03},
	journal = {Cold Spring Harbor Perspectives in Biology},
	author = {Pollard, Thomas D.},
	month = aug,
	year = {2016},
	pages = {a018226},
}

@article{khaitlina_intracellular_2014,
	title = {Intracellular transport based on actin polymerization},
	volume = {79},
	issn = {0006-2979, 1608-3040},
	url = {http://link.springer.com/10.1134/S0006297914090089},
	doi = {10.1134/S0006297914090089},
	language = {en},
	number = {9},
	urldate = {2025-12-03},
	journal = {Biochemistry (Moscow)},
	author = {Khaitlina, S. Yu.},
	month = sep,
	year = {2014},
	pages = {917--927},
}

@article{kanchanawong_nanoscale_2010,
	title = {Nanoscale architecture of integrin-based cell adhesions},
	volume = {468},
	copyright = {http://www.springer.com/tdm},
	issn = {0028-0836, 1476-4687},
	url = {https://www.nature.com/articles/nature09621},
	doi = {10.1038/nature09621},
	language = {en},
	number = {7323},
	urldate = {2025-12-03},
	journal = {Nature},
	author = {Kanchanawong, Pakorn and Shtengel, Gleb and Pasapera, Ana M. and Ramko, Ericka B. and Davidson, Michael W. and Hess, Harald F. and Waterman, Clare M.},
	month = nov,
	year = {2010},
	pages = {580--584},
}

@article{zaidel-bar_functional_2007,
	title = {Functional atlas of the integrin adhesome},
	volume = {9},
	copyright = {http://www.springer.com/tdm},
	issn = {1465-7392, 1476-4679},
	url = {https://www.nature.com/articles/ncb0807-858},
	doi = {10.1038/ncb0807-858},
	language = {en},
	number = {8},
	urldate = {2025-12-03},
	journal = {Nature Cell Biology},
	author = {Zaidel-Bar, Ronen and Itzkovitz, Shalev and Ma'ayan, Avi and Iyengar, Ravi and Geiger, Benjamin},
	month = aug,
	year = {2007},
	pages = {858--867},
}

@article{cronin_dynamics_2021,
	title = {Dynamics of the {Actin} {Cytoskeleton} at {Adhesion} {Complexes}},
	volume = {11},
	issn = {2079-7737},
	url = {https://www.mdpi.com/2079-7737/11/1/52},
	doi = {10.3390/biology11010052},
	language = {en},
	number = {1},
	urldate = {2025-12-03},
	journal = {Biology},
	author = {Cronin, Nicholas M. and DeMali, Kris A.},
	month = dec,
	year = {2021},
	pages = {52},
}

@article{roca-cusachs_integrin-dependent_2013,
	title = {Integrin-dependent force transmission to the extracellular matrix by alpha-actinin triggers adhesion maturation},
	volume = {110},
	issn = {0027-8424, 1091-6490},
	url = {https://pnas.org/doi/full/10.1073/pnas.1220723110},
	doi = {10.1073/pnas.1220723110},
	language = {en},
	number = {15},
	urldate = {2025-12-03},
	journal = {Proceedings of the National Academy of Sciences},
	author = {Roca-Cusachs, Pere and Del Rio, Armando and Puklin-Faucher, Eileen and Gauthier, Nils C. and Biais, Nicolas and Sheetz, Michael P.},
	month = apr,
	year = {2013},
}

@article{gupta_how_2021,
	title = {How adherens junctions move cells during collective migration},
	volume = {10},
	issn = {2732-432X},
	url = {https://facultyopinions.com/prime/reports/b/10/56/},
	doi = {10.12703/r/10-56},
	urldate = {2025-12-03},
	journal = {Faculty Reviews},
	author = {Gupta, Shafali and Yap, Alpha S},
	month = jun,
	year = {2021},
}

@article{pelham_cell_1997,
	title = {Cell locomotion and focal adhesions are regulated by substrate flexibility},
	volume = {94},
	issn = {0027-8424, 1091-6490},
	url = {https://pnas.org/doi/full/10.1073/pnas.94.25.13661},
	doi = {10.1073/pnas.94.25.13661},
	language = {en},
	number = {25},
	urldate = {2025-12-16},
	journal = {Proceedings of the National Academy of Sciences},
	author = {Pelham, Robert J. and Wang, Yu-li},
	month = dec,
	year = {1997},
	pages = {13661--13665},
}

@article{smith_stem_2018,
	title = {Stem {Cell} {Differentiation} is {Regulated} by {Extracellular} {Matrix} {Mechanics}},
	volume = {33},
	issn = {1548-9213, 1548-9221},
	url = {https://www.physiology.org/doi/10.1152/physiol.00026.2017},
	doi = {10.1152/physiol.00026.2017},
	language = {en},
	number = {1},
	urldate = {2025-12-16},
	journal = {Physiology},
	author = {Smith, Lucas R. and Cho, Sangkyun and Discher, Dennis E.},
	month = jan,
	year = {2018},
	pages = {16--25},
}

@article{sedzinski_polar_2011,
	title = {Polar actomyosin contractility destabilizes the position of the cytokinetic furrow},
	volume = {476},
	copyright = {http://www.springer.com/tdm},
	issn = {0028-0836, 1476-4687},
	url = {https://www.nature.com/articles/nature10286},
	doi = {10.1038/nature10286},
	language = {en},
	number = {7361},
	urldate = {2025-12-16},
	journal = {Nature},
	author = {Sedzinski, Jakub and Biro, Maté and Oswald, Annelie and Tinevez, Jean-Yves and Salbreux, Guillaume and Paluch, Ewa},
	month = aug,
	year = {2011},
	pages = {462--466},
}

@article{mcbeath_cell_2004,
	title = {Cell {Shape}, {Cytoskeletal} {Tension}, and {RhoA} {Regulate} {Stem} {Cell} {Lineage} {Commitment}},
	volume = {6},
	issn = {15345807},
	url = {https://linkinghub.elsevier.com/retrieve/pii/S1534580704000759},
	doi = {10.1016/S1534-5807(04)00075-9},
	language = {en},
	number = {4},
	urldate = {2025-12-16},
	journal = {Developmental Cell},
	author = {McBeath, Rowena and Pirone, Dana M and Nelson, Celeste M and Bhadriraju, Kiran and Chen, Christopher S},
	month = apr,
	year = {2004},
	pages = {483--495},
}

@article{kraning-rush_role_2011,
	title = {The role of the cytoskeleton in cellular force generation in {2D} and {3D} environments},
	volume = {8},
	issn = {1478-3975},
	url = {https://iopscience.iop.org/article/10.1088/1478-3975/8/1/015009},
	doi = {10.1088/1478-3975/8/1/015009},
	number = {1},
	urldate = {2025-12-16},
	journal = {Physical Biology},
	author = {Kraning-Rush, Casey M and Carey, Shawn P and Califano, Joseph P and Smith, Brooke N and Reinhart-King, Cynthia A},
	month = feb,
	year = {2011},
	pages = {015009},
}

@article{dembo_stresses_1999,
	title = {Stresses at the {Cell}-to-{Substrate} {Interface} during {Locomotion} of {Fibroblasts}},
	volume = {76},
	issn = {00063495},
	url = {https://linkinghub.elsevier.com/retrieve/pii/S0006349599773868},
	doi = {10.1016/S0006-3495(99)77386-8},
	language = {en},
	number = {4},
	urldate = {2025-12-16},
	journal = {Biophysical Journal},
	author = {Dembo, Micah and Wang, Yu-Li},
	month = apr,
	year = {1999},
	pages = {2307--2316},
}

@article{mohl_quantitative_2012,
	title = {Quantitative mapping of averaged focal adhesion dynamics in migrating cells by shape normalization},
	volume = {125},
	issn = {1477-9137, 0021-9533},
	url = {https://journals.biologists.com/jcs/article/125/1/155/32238/Quantitative-mapping-of-averaged-focal-adhesion},
	doi = {10.1242/jcs.090746},
	language = {en},
	number = {1},
	urldate = {2025-12-16},
	journal = {Journal of Cell Science},
	author = {Möhl, Christoph and Kirchgessner, Norbert and Schäfer, Claudia and Hoffmann, Bernd and Merkel, Rudolf},
	month = jan,
	year = {2012},
	pages = {155--165},
}

@article{plotnikov_force_2012,
	title = {Force {Fluctuations} within {Focal} {Adhesions} {Mediate} {ECM}-{Rigidity} {Sensing} to {Guide} {Directed} {Cell} {Migration}},
	volume = {151},
	issn = {00928674},
	url = {https://linkinghub.elsevier.com/retrieve/pii/S0092867412014195},
	doi = {10.1016/j.cell.2012.11.034},
	language = {en},
	number = {7},
	urldate = {2025-12-16},
	journal = {Cell},
	author = {Plotnikov, Sergey V. and Pasapera, Ana M. and Sabass, Benedikt and Waterman, Clare M.},
	month = dec,
	year = {2012},
	pages = {1513--1527},
}

@article{grashoff_measuring_2010,
	title = {Measuring mechanical tension across vinculin reveals regulation of focal adhesion dynamics},
	volume = {466},
	copyright = {http://www.springer.com/tdm},
	issn = {0028-0836, 1476-4687},
	url = {https://www.nature.com/articles/nature09198},
	doi = {10.1038/nature09198},
	language = {en},
	number = {7303},
	urldate = {2025-12-16},
	journal = {Nature},
	author = {Grashoff, Carsten and Hoffman, Brenton D. and Brenner, Michael D. and Zhou, Ruobo and Parsons, Maddy and Yang, Michael T. and McLean, Mark A. and Sligar, Stephen G. and Chen, Christopher S. and Ha, Taekjip and Schwartz, Martin A.},
	month = jul,
	year = {2010},
	pages = {263--266},
}

@article{banda_reference-free_2019,
	title = {Reference-{Free} {Traction} {Force} {Microscopy} {Platform} {Fabricated} via {Two}-{Photon} {Laser} {Scanning} {Lithography} {Enables} {Facile} {Measurement} of {Cell}-{Generated} {Forces}},
	volume = {11},
	copyright = {https://doi.org/10.15223/policy-029},
	issn = {1944-8244, 1944-8252},
	url = {https://pubs.acs.org/doi/10.1021/acsami.9b04362},
	doi = {10.1021/acsami.9b04362},
	language = {en},
	number = {20},
	urldate = {2025-12-16},
	journal = {ACS Applied Materials \& Interfaces},
	author = {Banda, Omar A. and Sabanayagam, Chandran R. and Slater, John H.},
	month = may,
	year = {2019},
	pages = {18233--18241},
}

@article{kollimada_biochemical_2021,
	title = {The biochemical composition of the actomyosin network sets the magnitude of cellular traction forces},
	volume = {32},
	issn = {1059-1524, 1939-4586},
	url = {https://www.molbiolcell.org/doi/10.1091/mbc.E21-03-0109},
	doi = {10.1091/mbc.E21-03-0109},
	language = {en},
	number = {18},
	urldate = {2025-12-16},
	journal = {Molecular Biology of the Cell},
	author = {Kollimada, Somanna and Senger, Fabrice and Vignaud, Timothée and Théry, Manuel and Blanchoin, Laurent and Kurzawa, Laëtitia},
	editor = {Dunn, Alexander},
	month = aug,
	year = {2021},
	pages = {1737--1748},
}

@article{hale_resolving_2009,
	title = {Resolving the {Role} of {Actoymyosin} {Contractility} in {Cell} {Microrheology}},
	volume = {4},
	issn = {1932-6203},
	url = {https://dx.plos.org/10.1371/journal.pone.0007054},
	doi = {10.1371/journal.pone.0007054},
	language = {en},
	number = {9},
	urldate = {2025-12-16},
	journal = {PLoS ONE},
	author = {Hale, Christopher M. and Sun, Sean X. and Wirtz, Denis},
	editor = {Kreplak, Laurent},
	month = sep,
	year = {2009},
	pages = {e7054},
}

@article{livne_self-assembled_2024,
	title = {Self-assembled active actomyosin gels spontaneously curve and wrinkle similar to biological cells and tissues},
	volume = {121},
	issn = {0027-8424, 1091-6490},
	url = {https://pnas.org/doi/10.1073/pnas.2309125121},
	doi = {10.1073/pnas.2309125121},
	language = {en},
	number = {2},
	urldate = {2025-12-16},
	journal = {Proceedings of the National Academy of Sciences},
	author = {Livne, Gefen and Gat, Shachar and Armon, Shahaf and Bernheim-Groswasser, Anne},
	month = jan,
	year = {2024},
	pages = {e2309125121},
}

@article{ideses_spontaneous_2018,
	title = {Spontaneous buckling of contractile poroelastic actomyosin sheets},
	volume = {9},
	issn = {2041-1723},
	url = {https://www.nature.com/articles/s41467-018-04829-x},
	doi = {10.1038/s41467-018-04829-x},
	language = {en},
	number = {1},
	urldate = {2025-12-16},
	journal = {Nature Communications},
	author = {Ideses, Y. and Erukhimovitch, V. and Brand, R. and Jourdain, D. and Hernandez, J. Salmeron and Gabinet, U. R. and Safran, S. A. and Kruse, K. and Bernheim-Groswasser, A.},
	month = jun,
	year = {2018},
	pages = {2461},
}

@article{alvarado_molecular_2013,
	title = {Molecular motors robustly drive active gels to a critically connected state},
	volume = {9},
	copyright = {2013 Springer Nature Limited},
	issn = {1745-2481},
	url = {https://www.nature.com/articles/nphys2715},
	doi = {10.1038/nphys2715},
	language = {en},
	number = {9},
	urldate = {2025-02-23},
	journal = {Nature Physics},
	author = {Alvarado, José and Sheinman, Michael and Sharma, Abhinav and MacKintosh, Fred C. and Koenderink, Gijsje H.},
	month = sep,
	year = {2013},
	note = {Publisher: Nature Publishing Group},
	pages = {591--597},
}

@article{alvarado_force_2017,
	title = {Force percolation of contractile active gels},
	volume = {13},
	issn = {1744-6848},
	url = {https://pubs.rsc.org/en/content/articlelanding/2017/sm/c7sm00834a},
	doi = {10.1039/C7SM00834A},
	language = {en},
	number = {34},
	urldate = {2025-02-23},
	journal = {Soft Matter},
	author = {Alvarado, José and Sheinman, Michael and Sharma, Abhinav and MacKintosh, Fred C. and Koenderink, Gijsje H.},
	month = aug,
	year = {2017},
	note = {Publisher: The Royal Society of Chemistry},
	pages = {5624--5644},
}

@article{amiri_intracellular_2023,
	title = {Intracellular tension sensor reveals mechanical anisotropy of the actin cytoskeleton},
	volume = {14},
	issn = {2041-1723},
	url = {https://www.nature.com/articles/s41467-023-43612-5},
	doi = {10.1038/s41467-023-43612-5},
	language = {en},
	number = {1},
	urldate = {2025-12-16},
	journal = {Nature Communications},
	author = {Amiri, Sorosh and Muresan, Camelia and Shang, Xingbo and Huet-Calderwood, Clotilde and Schwartz, Martin A. and Calderwood, David A. and Murrell, Michael},
	month = dec,
	year = {2023},
	pages = {8011},
}

@article{vogel_symmetry_2020,
	title = {Symmetry {Breaking} and {Emergence} of {Directional} {Flows} in {Minimal} {Actomyosin} {Cortices}},
	volume = {9},
	issn = {2073-4409},
	url = {https://www.mdpi.com/2073-4409/9/6/1432},
	doi = {10.3390/cells9061432},
	language = {en},
	number = {6},
	urldate = {2025-12-16},
	journal = {Cells},
	author = {Vogel, Sven K. and Wölfer, Christian and Ramirez-Diaz, Diego A. and Flassig, Robert J. and Sundmacher, Kai and Schwille, Petra},
	month = jun,
	year = {2020},
	pages = {1432},
}

@article{naganathan_active_2014,
	title = {Active torque generation by the actomyosin cell cortex drives left–right symmetry breaking},
	volume = {3},
	copyright = {http://creativecommons.org/licenses/by/4.0/},
	issn = {2050-084X},
	url = {https://elifesciences.org/articles/04165},
	doi = {10.7554/eLife.04165},
	language = {en},
	urldate = {2025-12-16},
	journal = {eLife},
	author = {Naganathan, Sundar Ram and Fürthauer, Sebastian and Nishikawa, Masatoshi and Jülicher, Frank and Grill, Stephan W},
	month = dec,
	year = {2014},
	pages = {e04165},
}

@article{ierushalmi_centering_2020,
	title = {Centering and symmetry breaking in confined contracting actomyosin networks},
	volume = {9},
	copyright = {http://creativecommons.org/licenses/by/4.0/},
	issn = {2050-084X},
	url = {https://elifesciences.org/articles/55368},
	doi = {10.7554/eLife.55368},
	language = {en},
	urldate = {2025-12-16},
	journal = {eLife},
	author = {Ierushalmi, Niv and Malik-Garbi, Maya and Manhart, Angelika and Abu Shah, Enas and Goode, Bruce L and Mogilner, Alex and Keren, Kinneret},
	month = apr,
	year = {2020},
	pages = {e55368},
}

@article{valencia_actin_2025,
	title = {Actin cytoskeleton protection by the formin-mediated safety valve},
	volume = {97},
	issn = {09550674},
	url = {https://linkinghub.elsevier.com/retrieve/pii/S0955067425001310},
	doi = {10.1016/j.ceb.2025.102593},
	language = {en},
	urldate = {2025-12-16},
	journal = {Current Opinion in Cell Biology},
	author = {Valencia, Fernando R. and Plotnikov, Sergey V.},
	month = dec,
	year = {2025},
	pages = {102593},
}

@article{valencia_force-dependent_2021,
	title = {Force-dependent activation of actin elongation factor {mDia1} protects the cytoskeleton from mechanical damage and promotes stress fiber repair},
	volume = {56},
	issn = {15345807},
	url = {https://linkinghub.elsevier.com/retrieve/pii/S1534580721008893},
	doi = {10.1016/j.devcel.2021.11.004},
	language = {en},
	number = {23},
	urldate = {2025-12-16},
	journal = {Developmental Cell},
	author = {Valencia, Fernando R. and Sandoval, Eduardo and Du, Joy and Iu, Ernest and Liu, Jian and Plotnikov, Sergey V.},
	month = dec,
	year = {2021},
	pages = {3288--3302.e5},
}

@article{smith_zyxin-mediated_2010,
	title = {A {Zyxin}-{Mediated} {Mechanism} for {Actin} {Stress} {Fiber} {Maintenance} and {Repair}},
	volume = {19},
	issn = {15345807},
	url = {https://linkinghub.elsevier.com/retrieve/pii/S1534580710003837},
	doi = {10.1016/j.devcel.2010.08.008},
	language = {en},
	number = {3},
	urldate = {2025-12-16},
	journal = {Developmental Cell},
	author = {Smith, Mark A. and Blankman, Elizabeth and Gardel, Margaret L. and Luettjohann, Laura and Waterman, Clare M. and Beckerle, Mary C.},
	month = sep,
	year = {2010},
	pages = {365--376},
}

@article{sala_stress_2021,
	title = {Stress fiber strain recognition by the {LIM} protein testin is cryptic and mediated by {RhoA}},
	volume = {32},
	issn = {1059-1524, 1939-4586},
	url = {https://www.molbiolcell.org/doi/10.1091/mbc.E21-03-0156},
	doi = {10.1091/mbc.E21-03-0156},
	language = {en},
	number = {18},
	urldate = {2025-12-16},
	journal = {Molecular Biology of the Cell},
	author = {Sala, Stefano and Oakes, Patrick W.},
	editor = {Dunn, Alex},
	month = aug,
	year = {2021},
	pages = {1758--1771},
}

@article{sakamoto_f-actin_2024,
	title = {F-actin architecture determines the conversion of chemical energy into mechanical work},
	volume = {15},
	issn = {2041-1723},
	url = {https://www.nature.com/articles/s41467-024-47593-x},
	doi = {10.1038/s41467-024-47593-x},
	language = {en},
	number = {1},
	urldate = {2025-12-16},
	journal = {Nature Communications},
	author = {Sakamoto, Ryota and Murrell, Michael P.},
	month = apr,
	year = {2024},
	pages = {3444},
}

@article{sakamoto_mechanical_2024,
	title = {Mechanical power is maximized during contractile ring-like formation in a biomimetic dividing cell model},
	volume = {15},
	issn = {2041-1723},
	url = {https://www.nature.com/articles/s41467-024-53228-y},
	doi = {10.1038/s41467-024-53228-y},
	language = {en},
	number = {1},
	urldate = {2025-12-16},
	journal = {Nature Communications},
	author = {Sakamoto, Ryota and Murrell, Michael P.},
	month = nov,
	year = {2024},
	pages = {9731},
}

\end{document}